\documentclass[12pt,a4paper]{article}
\pdfoutput=1
\usepackage{epsfig}
\usepackage{comment}
\usepackage{latexsym}
\usepackage{color}
\usepackage{amsmath}
\usepackage{hyperref}
\usepackage[english]{babel}

\newcommand{\mysquare}[0]{\raise-.2ex\hbox{{\Large$\Box$}}}
\def\lsim{\mathrel{\rlap {\raise.5ex\hbox{$ < $}}
{\lower.5ex\hbox{$\sim$}}}}
\def\gsim{\mathrel{\rlap {\raise.5ex\hbox{$ > $}}
{\lower.5ex\hbox{$\sim$}}}} \topmargin -1.5cm \textheight=22.5cm \textwidth=16.5cm
\catcode`\@=11
\newcount\hour
\newcount\minute
\newtoks\amorpm
\hour=\time\divide\hour by60 \minute=\time{\multiply\hour by60 \global\advance\minute by-\hour}
\edef\standardtime{{\ifnum\hour<12 \global\amorpm={am}%
        \else\global\amorpm={pm}\advance\hour by-12 \fi
        \ifnum\hour=0 \hour=12 \fi
        \number\hour:\ifnum\minute<10 0\fi\number\minute\the\amorpm}}
\edef\militarytime{\number\hour:\ifnum\minute<10 0\fi\number\minute}
\def\draftlabel#1{{\@bsphack\if@filesw {\let\thepage\relax
   \xdef\@gtempa{\write\@auxout{\string
      \newlabel{#1}{{\@currentlabel}{\thepage}}}}}\@gtempa
   \if@nobreak \ifvmode\nobreak\fi\fi\fi\@esphack}
        \gdef\@eqnlabel{#1}}
\def\@eqnlabel{}
\def\@vacuum{}
\def\draftmarginnote#1{\marginpar{\raggedright\scriptsize\tt#1}}
\def\draft{\oddsidemargin -.2truein
        \def\@oddfoot{\sl preliminary draft \hfil
        \rm\thepage\hfil\sl\today\quad\militarytime}
        \let\@evenfoot\@oddfoot \overfullrule 3pt
        \let\label=\draftlabel
        \let\marginnote=\draftmarginnote
   \def\@eqnnum{(\theequation)\rlap{\k

 ern\marginparsep\tt\@eqnlabel}%
\global\let\@eqnlabel\@vacuum}  }

\newcommand{\be}[0]{\begin{equation}}
\newcommand{\ee}[0]{\end{equation}}
\newcommand{\ba}[0]{\begin{eqnarray}}
\newcommand{\ea}[0]{\end{eqnarray}}

\def\bs{\begin{subequations}}
\def\es{\end{subequations}}

\def\thebibliography#1{%
\vskip 0.5cm \centerline{\bf \Large References}
\list{%
[\arabic{enumi}]}{\settowidth\labelwidth{[#1]} \leftmargin\labelwidth \advance\leftmargin\labelsep
\usecounter{enumi}}
\def\newblock{\hskip .11em plus .33em minus .07em}
\sloppy\clubpenalty4000\widowpenalty4000 \sfcode`\.=1000\relax}

\renewcommand{\theequation}{\arabic{section}.\arabic{equation}}

\renewcommand{\section}{\setcounter{equation}{0}\@startsection
{section}{1}{0mm}{-\baselineskip}{0.5\baselineskip} {\normalfont\Large\bfseries}}

\renewcommand{\subsection}{\@startsection
{subsection}{2}{0mm}{-\baselineskip}{0.5\baselineskip} {\normalfont\large\bfseries}}

\renewcommand{\subsubsection}{\@startsection
{subsubsection}{3}{0mm}{-\baselineskip}{0.5\baselineskip} {\normalfont\normalsize\slshape}}

\usepackage{amssymb,amsfonts}
\usepackage{graphicx}
\usepackage{cite}

\newcommand{\bea}{\begin{eqnarray}}
\newcommand{\eea}{\end{eqnarray}}
\newcommand{\dis}{\displaystyle}

\newcommand{\Co}{\mathbb{C}}
\newcommand{\Z}{\mathbb{Z}}
\newcommand{\Ka}{K{\"a}hler }
\renewcommand{\O}{{\cal O}}

\newcommand{\abs}{|}

\newcommand{\ie}{{\em i.e. }}
\newcommand{\where}{\mbox{where}}

\newcommand{\when}{\mbox{when}}
\renewcommand{\and}{\mbox{and}}

\newcommand{\F}{{\cal F}}
\newcommand{\N}{{\cal N}}
\newcommand{\A}{{\cal A}}
\newcommand{\B}{{\cal B}}
\newcommand{\K}{{\cal K}}
\newcommand{\T}{{\cal T}}
\newcommand{\X}{{\cal X}}
\newcommand{\U}{{\cal U}}
\newcommand{\M}{{\cal M}}
\renewcommand{\S}{{\cal S}}
\newcommand{\Y}{{\cal Y}}

\renewcommand{\b}{\bar}
\newcommand{\h}{\hat}
\renewcommand{\v}{\vec}
\renewcommand{\t}{\tilde}
\newcommand{\rmI}{{\rm I}}

\def\cf{{\cal F}}

\def\cs{{\cal S}}
\def\ct{{\cal T}}
\def\cu{{\cal U}}

\def\cw{{\cal W}}

\def\nnR{\nonumber\\}

\newcommand{\demi}{\frac{1}{2}}

\def\lst{\left|}
\def\rst{\right|}
\def\tdt{\tilde{\ct}}

\begin{document}
\begin{titlepage}
\begin{flushright}
LPTENS--11/07, CPHT--RR003.0211, February 2011
\end{flushright}

\vspace{-2mm}

\begin{centering}

{\bf \Large MASSLESS D-STRINGS AND MODULI\\
\vspace{3mm}
STABILIZATION IN TYPE I COSMOLOGY}

\vspace{8mm}
 {\Large John Estes$^{1,2}$, Lihui Liu$^{3}$ and Herv\'e Partouche$^3$}

\vspace{4mm}

$^1$ Instituut voor Theoretische Fysica, Katholieke Universiteit Leuven, \\
Celestijnenlaan 200D B-3001 Leuven, Belgium\\
{\em johnalondestes@gmail.com}

\vspace{2mm}

$^2$ Laboratoire de Physique Th\'eorique,
Ecole Normale Sup\'erieure,$^\dag$ \\
24 rue Lhomond, F--75231 Paris cedex 05, France

\vspace{2mm}

$^3$ Centre de Physique Th\'eorique, Ecole Polytechnique,$^\ddag$
\\
F--91128 Palaiseau cedex, France\\
{\em lihui.liu@cpht.polytechnique.fr}\\
{\em herve.partouche@cpht.polytechnique.fr}
\vspace{8mm}

{\bf\Large Abstract}

\end{centering}
\vspace{4mm}

\noindent
We consider the cosmological evolution induced by the free energy $F$ of a gas of maximally supersymmetric heterotic strings at finite temperature and weak coupling in dimension $D\ge 4$. We show that $F$, which plays the role of an effective potential, has minima associated to enhanced gauge symmetries, where all internal moduli can be attracted and dynamically stabilized. Using the fact that the heterotic/type I S-duality remains valid at finite temperature and can be applied at each instant of a quasi-static evolution, we find in the dual type I cosmology that all internal NS-NS and RR moduli in the closed string sector and the Wilson lines in the open string sector can be stabilized.  For the special case of $D=6$, the internal volume modulus remains a flat direction, while the dilaton is stabilized.  An essential role is played by light D-string modes wrapping the internal manifold and whose contribution to the free energy cannot be omitted, even when the type I string is at weak coupling. As a result, the order of magnitude of the internal radii expectation values on the type I side is $\sqrt{\lambda_\rmI \, \alpha'}$, where  $\lambda_\rmI$ is the ten-dimensional string coupling.  The non-perturbative corrections to the type I free energy can alternatively be described as effects of ``thermal E1-instantons'', whose worldsheets wrap the compact Euclidean time cycle.


\vspace{3pt} \vfill \hrule width 6.7cm \vskip.1mm{\small \small \small
\noindent
 $^\dag$\ Unit{\'e} mixte  du CNRS et de l'Ecole Normale Sup{\'e}rieure associ\'ee \`a
l'Universit\'e Pierre et Marie Curie (Paris 6), UMR 8549.\\
$^\ddag$\ Unit{\'e} mixte du CNRS et de l'Ecole Polytechnique,
UMR 7644.}

\end{titlepage}
\newpage
\setcounter{footnote}{0}
\renewcommand{\thefootnote}{\arabic{footnote}}
 \setlength{\baselineskip}{.7cm} \setlength{\parskip}{.2cm}

\setcounter{section}{0}


\section{Introduction}
\label{intro}
The $SO(32)$ heterotic and type I strings are dual perturbative descriptions of the same underlying theory \cite{Polchinski:1995df}. This is easily observed at the level of the low energy effective actions, which are equivalent after certain field redefinitions.  This follows from the fact that short massless supermultiplets have protected masses and that $\N_{10}=1$ supergravity coupled to ten dimensional super Yang-Mills theory with given gauge group is unique at the two-derivative level.  One interesting facet of the equivalence is that in ten dimensions, the heterotic and type I string couplings are inverse to one another, and thus one has the opportunity to uncover strong coupling effects.  In dimension $D\ge 7$ ($D\le 5$), this leads to a strong-weak (weak-weak) duality, while for $D=6$, string couplings and internal volumes are interchanged \cite{allD}.

In the literature, most of the applications of string dualities have been based on BPS states and therefore restricted to models where supersymmetry is preserved in static universes. In general, extending these ideas to non-supersymmetric cases (see \cite{Blum:1997gw} for some examples) and cosmological evolutions is difficult. However, such a project can still be addressed within the context of no-scale models \cite{Noscale}. The latter are defined at the classical level by backgrounds associated to vanishing minima of a scalar potential, which admit a flat direction parameterized by the scale of spontaneous supersymmetry breaking.  The non-trivial vacuum energy, which arises at the quantum level, backreacts on the flat and originally static universe, and induces a quasi-static time evolution in the background fields \cite{Kounnas:2007hb}.

To be specific, start with a dual pair of supersymmetric heterotic and type I models.  As follows from the adiabatic argument of \cite{Vafa:1995gm}, one may implement on both sides a spontaneous breaking of supersymmetry, thus giving a new dual pair.  For example, if the heterotic theory is in a perturbative regime and the spontaneous breaking at the classical level is compatible with flat Minkowski space, the cosmological evolution induced at the one-loop level can be reinterpreted in the dual type I regime.  In this paper, we spontaneously break supersymmetry by considering the models at finite temperature.  This can be implemented at the level of the two dimensional CFT by compactifying the Euclidean time on a circle, whose boundary conditions depend on the fermion number \cite{Kounnas:1989dk}.  In this case, the one-loop heterotic effective potential discussed above is nothing but the free energy of a perfect gas of supersymmetric strings.  Applying the heterotic/type I duality, we find the existence of novel contributions to the type I effective potential coming from light D-strings.  Despite being non-perturbative, these corrections have a large impact on the cosmological evolution, as well as on the low energy spectrum of the theory, even at weak type I string coupling.

A second method to spontaneously break supersymmetry is by introducing ``geometric'' fluxes along internal cycles \cite{Rohm}.  When the R-symmetry charge associated to the flux is the fermion number, this method is related to the finite temperature case by a double Wick rotation.  In this paper, we only explore the thermal breaking for simplicity and clarity, as most of our results have a direct generalization to the second case.  In realistic situations, one must include zero temperature spontaneous supersymmetry breaking before switching on finite temperature. In this case, a general picture arises, where the induced cosmology can be divided into different stages.  In the Hagedorn era, where the temperature $T$ is close to the string scale $M_{\rm s}$, a phase transition between pre- and post-big bang evolutions takes place. It can be described along the lines of Refs \cite{Hage,Florakis:2010is} at the level of the two dimensional CFT and is both free of initial singularity and consistent with perturbation theory. As the temperature drops, the cosmology induced by the one-loop effective potential can be trusted until infrared effects become relevant, such as in the cases of radiative breaking or confining gauge groups. For example, in standard GUT scenarios, this defines intermediate eras where the temperature evolves in either of the ranges $M_{\rm s}>T> \Lambda_{\rm GUT}$ or $\Lambda_{\rm GUT}>T>M_{\rm EW}$, where  $\Lambda_{\rm GUT}$ and $M_{\rm EW}$ are the GUT and electroweak scales \cite{cosmoA,cosmoB,cosmoC,cosmoreview}. These intermediate eras are connected by a phase transition where the dynamics responsible for the breaking of the GUT group must be taken account.  After the electroweak phase transition, the conventional history of the universe follows with the hadronic, leptonic and nucleosynthesis eras...

One feature of the above Hagedorn and intermediate eras is the possibility to stabilize internal moduli \cite{cosmoB,cosmoD,cosmoDb}. This is an important issue since current observations of the gravitational force place lower limits on scalar masses (see for example \cite{Adelberger:2003zx}).  Many approaches address this question by considering compactification spaces where (geometrical or non-geometrical) internal fluxes are switched on at the outset, while preserving some amount of supersymmetry \cite{geneCY}.  This leads to a partial stabilization since flat directions always persist in such models, at least at the perturbative level.  However, we would like to stress that once supersymmetry is broken, flat directions are generically lifted in string theory. This was considered long ago in non-supersymmetric heterotic string backgrounds, such as the $SO(16)\times SO(16)$ tachyon free theory toroidally compactified \cite{min}. However, minimization of the moduli-dependent ``cosmological constant" generated by loop corrections in such models leads to an unacceptably large vacuum energy at the minima, since supersymmetry is explicitly broken at the string scale. In \cite{Patil:2004zp}, it was realized that a gas of string modes, which carry both winding and momenta, generate a free energy that enables stabilization of radii moduli.  Upon introducing a zero temperature spontaneous breaking of supersymmetry at the string tree level, it was shown in \cite{cosmoC,cosmoD,cosmoDb,cosmoreview,Angelantonj:2006ut} that this effect also has a quantum version, with the thermal gas and free energy replaced by virtual strings which induce an effective potential\footnote{In Refs \cite{Kofman:2004yc}, the effect of the  Coleman-Weinberg effective potential is  explicitly subtracted in order to isolate the backreaction on the moduli arising from particle production near extra massless species points. To be substantial, this mechanism supposes the moduli already  have non-trivial motions at tree level. Since the no-scale models are based on classical static backgrounds, the moduli velocities occur as backreactions of the one-loop effective potential and particle production is higher order in perturbation theory.}.  An advantage of this type of stabilization is that during the intermediate eras, the induced masses are not constant. Instead, they follow the time-evolution of the temperature $T(t)$ and supersymmetry breaking modulus $M(t)$, which drop proportionally. It is only after the electroweak phase transition that $M(t)$ is stabilized and that the induced moduli masses become constant.  As a result, the energy of the moduli with time-dependent masses is diluted during the intermediate eras, and
the cosmological moduli problem \cite{cosmomodprob}\footnote{A simplified statement of this problem is that the energy of scalars with constant masses dilutes slower than the thermal energy of radiation, and so heavy scalars tend to dominate at late times, which can cause problems for nucleosynthesis.  This may be fixed by requiring the heavy scalars to be unstable so that their fluctuations eventually decay, thereby reheating space-time.  However, the reheating process creates extra entropy and one can run into problems with baryogensis.} is avoided. Moreover, the decrease of $M(t)\propto T(t)$ during the intermediate eras gives a dynamical explanation of the hierarchy between the supersymmetry breaking scale and the string scale,  $M\ll M_s$.

This above dynamical moduli stabilization relies on the existence of perturbative states in the string spectrum, whose masses are determined by the expectation value of the moduli and vanish at the stabilization points. For instance, in toroidal or orbifold compactifications of the heterotic string, if the radius $R_i$ of some factorized internal circle is not participating in the spontaneous breaking of supersymmetry, it can be attracted to the self dual point $R_i=1$  associated to an enhanced $SU(2)$ level one Kac Moody algebra. Another simple example can be realized in type II superstring, when the internal circle is used to spontaneously break the supersymmetries generated by the right-moving sector via the Scherck Schwarz mechanism. In this case, $R_i$ can be stabilized at the fermionic point $R_i=1/\sqrt{2}$ corresponding to a Kac Moody level two $SU(2)$ extension \cite{Hage}.  However, since this type II setup is intrinsically left/right asymmetric, it cannot be extended to orientifold models in a straightforward way.  Thus, the purpose of the present work is to infer how the internal moduli in type I no-scale models are stabilized by using our knowledge of the dual heterotic picture.  As said before, we consider only thermal effects, as this is sufficient to uncover the mechanism.  More specifically, using heterotic/type I duality at finite temperature, we infer the existence of non-pertrubative contributions to the thermal free energy of type I superstrings.  These contributions are due to light, or even massless, D-strings which wrap the internal cycles and participate to the dynamical stabilization of  all the internal moduli, including those in the RR sector and the Wilson lines.

We derive in section \ref{naiveTI} the free energy of a gas of weakly coupled  perturbative states in type I superstring, in the simple case where the internal space is a factorized torus. We describe the induced cosmological evolution and find the radii moduli are flat directions of the thermal potential.  In section \ref{dual}, using the dual heterotic model at weak coupling, we correct this naive analysis by taking into account contributions of non-perturbative states to the free energy. In particular, D-strings modes are found to be light when the radii are in a neighborhood of $\sqrt{\lambda_\rmI}$, where $\lambda_\rmI\gg 1$ is the ten dimensional type I string coupling. They produce local minima of the thermal potential which are responsible for the stabilization of the radii at  $\sqrt{\lambda_\rmI}$. In type I, this dynamical effect occurs at strong (weak) coupling when $D\ge 7$ ($D\le 6$).  However, since the BPS masses of the light D1-branes are protected by supersymmetry, our results are also valid at small string coupling for $D\ge 7$.  In section \ref{E1}, we reexamine the form of the corrections to the free energy along the lines of \cite{Bachas:1997mc}, and interpret the non-perturbative contributions as arising from ``thermal E1-instantons''.  What is meant by this is that the Euclidean worldsheets of the D1-branes wrap the Euclidean time circle.
In section \ref{stab}, we generalize our results : The one-loop heterotic free energy is computed, with all of the internal moduli taken into account. We find that at certain points in moduli space, all scalars, except the dilaton, may be stabilized for $D\ge 4$.\footnote{Additionally, for $D \geq 5$ the dilaton approaches a constant finite value at late times and the cosmological evolution is radiation dominated. For $D=4$, the dilaton decreases logarithmically with cosmological time and the coherent motion of all moduli is such that the metric evolution is that of a radiation dominated universe, $H^2\propto 1/a^4$. However, non-perturbative effects from NS5 or D5-branes in the heterotic or type I theories should be taken into account in four dimensions and may play a role in stabilizing the dilaton.}   On the dual type I side, the non-perturbative effects induce a stabilization of the internal NS-NS and RR moduli in the closed string sector, and the Wilson lines in the open string sector.  For the special case of $D=6$, the internal volume modulus remains a flat direction, while the dilaton is stabilized at a small value.  In section \ref{example}, we give explicit examples of loci in moduli space where only the flat direction of the dilaton survives. Section \ref{conclu} is devoted to our conclusions and perspectives.


\section{Naive perturbative type I thermal cosmology}
\label{naiveTI}

In this   section, we derive the cosmology induced by thermal effects in the purely perturbative type I superstring theory toroidally compactified down to $D\ge 3$ dimensions. We shall see in the next   section how light solitonic states correct this picture in a drastic way. In the following, quantities  are denoted in the type I context with subscripts I and, throughout this paper, ``hatted'' (``un-hatted") ones  are referring to  the string (Einstein) frame. Finite temperature $\h T_{\rm I}$ is implemented by considering an Euclidean time of period $\h \beta_{\rm I}=2\pi R_{{\rm I}0} = 1/\h T_{\rm I}$, and coupling the associated  $S^1(R_{{\rm I}0})$ lattice of zero modes  to the fermion number. We restrict for the moment our study to the case of a factorized internal space $\prod_{i=D}^9 S^1(R_{{\rm I}i})$ and analyze the dynamics of the radii $R_{{\rm I}i}$.

Working in a perturbative regime, there are four contributions to the Euclidean one-loop partition function needed to express the free energy density, namely the torus, Klein-bottle, annulus and  M\"obius strip vacuum-to-vacuum amplitudes $\T$, $\K$, $\A$ and $\M$. In units where $\alpha'=1$,
a little work yields (see the appendix),
\be
\label{ZT}
\T= {\h \beta_{\rm I} \hat V_{\rm I}\over \h\beta_{\rm I}^D} \Bigg\{ s_0^2\, c_D+ \!\!\!\sum_{\scriptsize \substack{A\ge 0, \, \b A\ge 0, \,  \v m, \, \v n\\ A-\b A=\v m\cdot \v n\\ (A,\v m,\v n)\neq (0,\v 0,\v 0)}} \!\!\!  \!\!\!s_As_{\b A}\, G\bigg(2\pi R_{{\rm I}0}\Big[4A+\sum_{i=D}^9\Big({m_i\over R_{{\rm I}i}}-n^iR_{{\rm I}i}\Big)^2\Big]^{1\over 2}\bigg)\Bigg\},
\ee
where $\h V_{\rm I}$ is the regularized volume of the $(D-1)$-dimensional space, $c_D$ is Stefan's constant for radiation in dimension $D$ and the function $G$ is defined in terms of a modified Bessel function of the second kind, $K_{D\over 2}(x)$ :
\be
\label{cg}
c_D={\Gamma({D\over 2})\over \pi^{D\over 2}}\sum_{\t k_0}{1\over \abs 2\t k_0+1\abs ^D}\; , \qquad G(x)= 2\sum_{\t k_0} \left({x\over 2\pi \abs 2\t k _0+1\abs }\right)^{D\over 2} K_{D\over 2}\big(x\, \abs 2 \t k_0+1\abs\big).
\ee
The integer $s_A$ ($s_{\b A}$) counts the degeneracy at oscillator level $A$ ($\b A$) on the left (right)-moving side of the worldsheet, while $m_i$ ($n^i$) labels the momentum (winding) number along the $i$-th cycle of the internal torus\footnote{Note that the condition $A-\b A=\v m\cdot \v n$ provides the level matching.}.   In (\ref{ZT}), the first term in the braces is the contribution of the massless modes, with quantum numbers $(A,\v m,\v n)=(0,\v 0,\v 0)$ and associated to the $\N_{10}=1$ supergravity multiplet in ten dimensions. The Klein-bottle contribution $\K$ vanishes.
The annulus plus M\"obius amplitude takes in a similar way the form
\be
\label{ZA}
\A+\M= {\h \beta_{\rm I} \hat V_{\rm I}\over \h\beta_{\rm I}^D} \Bigg\{  {N^2-N\over 2}\,  s_0\, c_D+\!\!\!\!\sum_{\scriptsize \substack{A\ge 0, \,  \v m\\ (A,\v m)\neq (0,\v 0) }}\!\!  \!\!\!{N^2-(-1)^A N\over 2}\, s_A\, G\bigg(2\pi R_{{\rm I}0}\Big[A+\sum_{i=D}^9\Big({m_i\over R_{{\rm I}i}}\Big)^2\Big]^{1\over 2}\bigg)\Bigg\},
\ee
where $N=32$ and the first term is associated to the $\N_{10}=1$ $SO(32)$ super-vector multiplet in ten dimensions.  The partition function is given by the sum $Z_{\rm I}=\T+\K+\A+\M$. At high temperatures, it becomes ill-defined.  Examining $\T$, one finds that winding modes along the Euclidean time circle become tachyonic when $R_{{\rm I}0}< R_{{\rm I}{\rm H}}$, where $R_{{\rm I}{\rm H}}=\sqrt{2}$ is the Hagedorn radius.  This divergence of $Z_{\rm I}$ is not a sickness of the theory, but rather the signal of a phase transition \cite{AtickWitten}.  From now on, we restrict ourselves to temperatures below $\h T_{{\rm I}{\rm H}} \equiv 1/(2\pi R_{{\rm I}{\rm H}})$.

The free energy density is defined in terms of the partition function as $\h\F_{\rm I}=-Z_{\rm I}/(\h \beta_{\rm I}\h V_{\rm I})$. It is expressed in terms of the $G$-function, whose arguments are the ratios of the spectrum masses to the temperature. Since
\be
\label{G}
G(x)=c_D-{c_{D-2}\over 4\pi}\, x^2+\O(x^4) \; \;\when \; \;x\simeq 0\; , \quad G(x)\sim \left({x\over 2\pi}\right)^{D-1\over 2} e^{-x} \; \;\when  \; \; x\gg 1,
\ee
the dominant contribution at low temperature (compared to the string scale) arises from the first terms of (\ref{ZT}) and (\ref{ZA}) and corresponds to the free energy density of thermal radiation,
\be
\h\F_{\rm I} = -  \left( s_0^2+{N^2-N\over 2}\, s_0\right) c_D \,  \h T_{\rm I}^D + \cdots .
\ee
However, if some $R_{{\rm I}i}$ is large (small) enough, $R_{{\rm I}i} > 2 \pi R_{{\rm I}0}$ ($R_{{\rm I}i} < 1/(2 \pi R_{{\rm I}0})$), pure Kaluza-Klein (winding) modes yield additional terms of the same order.
The contributions associated to the remaining states are exponentially suppressed.

It is straightforward to apply the techniques introduced in \cite{cosmoD,cosmoDb} for closed strings to study the backreaction of the type I free energy on the originally static background. For arbitrary initial conditions at the exit of the Hagedorn era, one finds that the system is attracted to a radiation dominated evolution, where all internal radii and the dilaton are frozen at non-specific values depending on the initial data.  Quantitatively, the final constant values of the $R_{{\rm I}i}$'s sit in the range
\be
\label{range}
{1\over 2\pi R_{{\rm I}0}}< R_{{\rm I}i}<2\pi R_{{\rm I}0}\; , \quad i=D,\dots, 9,
\ee
where $R_{{\rm I}0}$ is increasing with time, corresponding to an expanding and cooling universe.
Actually, if at some time $t$ a radius $R_{{\rm I}j}$ is outside this range, we find $R_{{\rm I}j}(t)$ and $R_{{\rm I}0}(t)$ always evolve so that the condition (\ref{range}) is finally satisfied, after which the evolution of $R_{{\rm I}j}$ comes to a halt.  This may be seen by examining the force on the modulus $\mu_j = \ln (2\pi R_{{\rm I}0}/R_{{\rm I}j})$ (or $\ln (2\pi R_{{\rm I}0}R_{{\rm I}j})$) \cite{cosmoD,cosmoDb}.

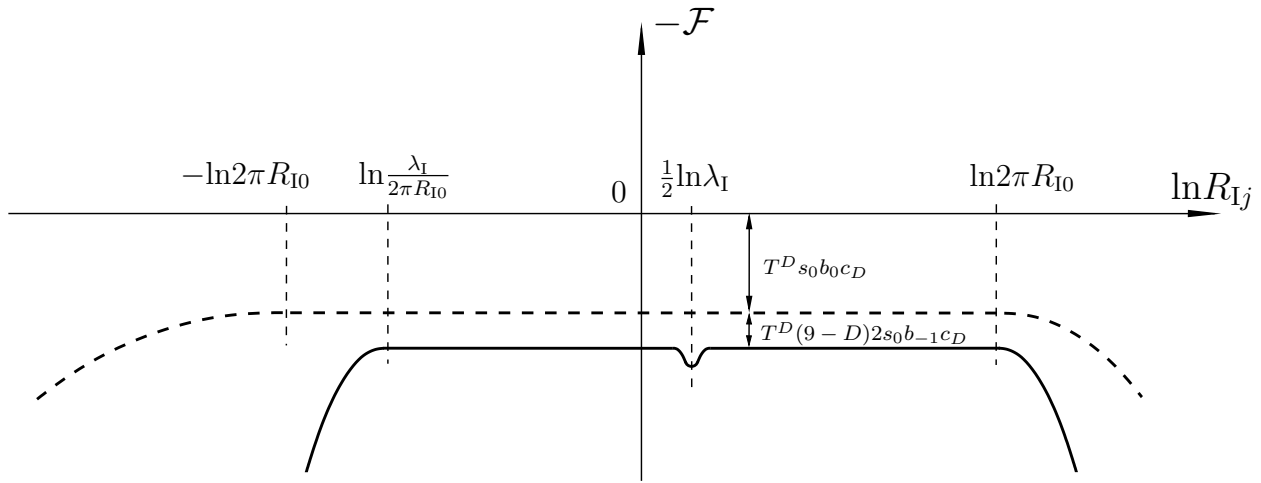
\begin{figure}

\begin{picture}(0,0)%
\includegraphics{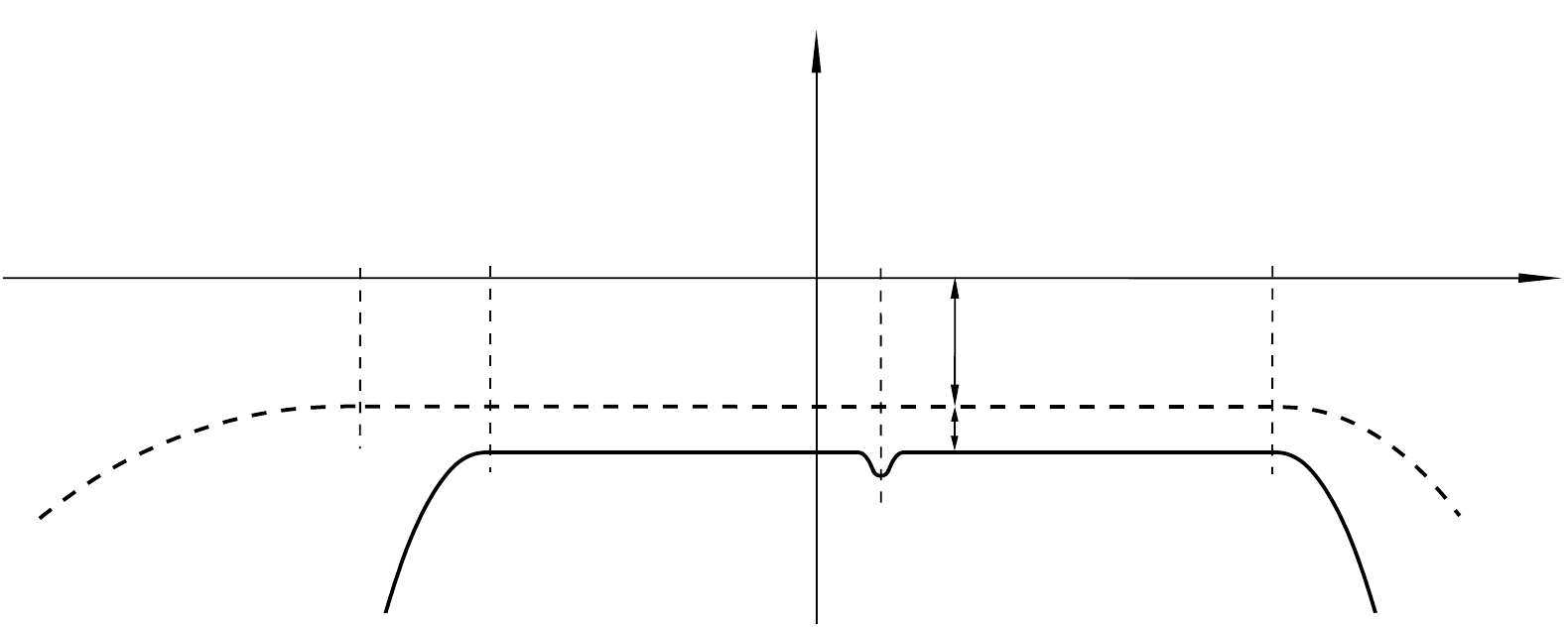}%
\end{picture}%
%
%
\setlength{\unitlength}{2368sp}%
\begingroup\makeatletter\ifx\SetFigFont\undefined%
\gdef\SetFigFont#1#2#3#4#5{%
  \reset@font\fontsize{#1}{#2pt}%
  \fontfamily{#3}\fontseries{#4}\fontshape{#5}%
  \selectfont}%
\fi\endgroup%
\begin{picture}(12613,5038)(40,-5888)
\put(7886,-3706){\makebox(0,0)[lb]{\smash{{\SetFigFont{8}{9.6}{\rmdefault}{\mddefault}{\updefault}{\color[rgb]{0,0,0}$T^Ds_0b_0c_D$}%
}}}}
\put(7858,-4397){\makebox(0,0)[lb]{\smash{{\SetFigFont{8}{9.6}{\rmdefault}{\mddefault}{\updefault}{\color[rgb]{0,0,0}$T^D(9-D)2s_0b_{-1}c_D$}%
}}}}
\put(6736,-1201){\makebox(0,0)[lb]{\smash{{\SetFigFont{14}{16.8}{\rmdefault}{\mddefault}{\updefault}{\color[rgb]{0,0,0}$-\cal{F}$}%
}}}}
\put(6309,-2951){\makebox(0,0)[lb]{\smash{{\SetFigFont{12}{14.4}{\rmdefault}{\mddefault}{\updefault}{\color[rgb]{0,0,0}$0$}%
}}}}
\put(1846,-2761){\makebox(0,0)[lb]{\smash{{\SetFigFont{12}{14.4}{\rmdefault}{\mddefault}{\updefault}{\color[rgb]{0,0,0}$-{\rm ln} 2\pi R_{{\rm I}0}$}%
}}}}
\put(3691,-2761){\makebox(0,0)[lb]{\smash{{\SetFigFont{12}{14.4}{\rmdefault}{\mddefault}{\updefault}{\color[rgb]{0,0,0}${\rm ln}{\lambda_{\rm I}\over 2\pi R_{{\rm I}0}}$}%
}}}}
\put(6796,-2791){\makebox(0,0)[lb]{\smash{{\SetFigFont{12}{14.4}{\rmdefault}{\mddefault}{\updefault}{\color[rgb]{0,0,0}${1\over 2}{\rm ln}\lambda_{\rm I}$}%
}}}}
\put(10021,-2791){\makebox(0,0)[lb]{\smash{{\SetFigFont{12}{14.4}{\rmdefault}{\mddefault}{\updefault}{\color[rgb]{0,0,0}${\rm ln}2\pi R_{{\rm I}0}$}%
}}}}
\put(12106,-2926){\makebox(0,0)[lb]{\smash{{\SetFigFont{14}{16.8}{\rmdefault}{\mddefault}{\updefault}{\color[rgb]{0,0,0}${\rm ln}R_{{\rm I}j}$}%
}}}}
\end{picture}%

\caption{\small Thermal effective potential (in Einstein frame) for $R_{{\rm I}j}$, when all other internal radii satisfy (\ref{range}). The dashed curve takes only into account the perturbative type I states. The solid one is obtained by heterotic/type I S-duality and receives corrections from light D-string modes.}
\label{pot}
\end{figure}

A difference compared to the type II and heterotic string cases, is that the open string sector is not invariant under T-duality, $R_{{\rm I}i}\to 1/R_{{\rm I}i}$ (for any $i$), due to a lack of winding quantum numbers in the open sector. For instance, for arbitrary $R_{{\rm I}j}$ (for a given $j$), while the other radii satisfy (\ref{range}), the effective potential for $R_{{\rm I}j}$, which is exactly the free energy density, simplifies to
\be
\label{FI}
\begin{array}{l}
\!\!\dis \h\F_{\rm I}= -\h T_{\rm I}^D\Bigg\{\!\!\left( s_0^2+{N^2-N\over 2}\, s_0\right)\bigg[c_D+\sum_{m_{j}\neq 0}G\Big(2\pi R_{{\rm I}0}{\abs m_{j}\abs \over R_{{\rm I}j}}\Big)\bigg]+\O(e^{-2\pi R_{{\rm I}0}})\Bigg\},\;\;\;\, 2\pi R_{{\rm I}0}<R_{{\rm I}j}\\
\!\!\dis\h \F_{\rm I}= -\h T_{\rm I}^D\Bigg\{\!\!\left( s_0^2+{N^2-N\over 2}\, s_0\right) c_D+\O(e^{-2\pi R_{{\rm I}0}})\Bigg\} ,\,\hspace{3.15cm} {1\over 2\pi R_{{\rm I}0}}<R_{{\rm I}j}<2\pi R_{{\rm I}0}\\
\!\!\dis\h \F_{\rm I}= -\h T_{\rm I}^D\Bigg\{{N^2-N\over 2}\, s_0\, c_D+s_0^2\, \bigg[c_D+\sum_{n_{j}\neq 0}G\Big(2\pi R_{{\rm I}0}\abs n_{j}\abs R_{{\rm I}j}\Big)\bigg]+\O(e^{-2\pi R_{{\rm I}0}})\Bigg\},  R_{{\rm I}j}<{1\over 2\pi R_{{\rm I}0}}
\end{array}
\ee
and is shown in figure \ref{pot}, in Einstein frame. When $R_{{\rm I}j}<1$, the theory is actually better understood in the T-dual type I' picture obtained by inverting $R_{{\rm I}j}$. More importantly, there is no local minimum of the free energy density where $R_{{\rm I}j}$ (as well as the $R_{{\rm I}i}$'s) can be attracted and stabilized. This is contrary to the heterotic case, where enhanced symmetry points exist and imply a local increase of the number of massless states. However, we shall find that the above purely perturbative analysis is missing important contributions from massless solitons.

\section{Heterotic/type I cosmological duality}
\label{dual}

Given that heterotic and type I theories at zero temperature are S-dual in  ten dimensions, it is a simple but non-trivial fact that they remain S-dual at finite temperature. Technically, the backgrounds used to analyze the thermal ensembles are freely acting orbifolds, obtained by modding out with $(-1)^F\delta_0$, where $\delta_0$ is an order-two shift along the Euclidean time circle and $F$ is the fermion number.  Using the ``adiabatic argument'' of \cite{Vafa:1995gm}, after such a free action, the two theories remain dual. Since the cosmological evolutions we study are quasi-static, it is valid to apply at each time an S-duality transformation on the heterotic side, in order to derive non-perturbative contributions to the type I free energy and its resulting backreaction.

\vspace{-.3cm}
\subsection*{\it S-dual $SO(32)$ heterotic string}
\vspace{-.3cm}
Let us apply this point of view to the type I background considered in   section \ref{naiveTI}.  The dual theory is the $SO(32)$ heterotic string compactified on $\prod_{i=D}^9 S^1(R_{{\rm h}i})$, where we use the subscript h to denote heterotic quantities.  As in the type I case, the partition function is only well defined when the temperature $\h T_{\rm h}=1/\h \beta_{\rm h}=1/(2\pi R_{{\rm h}0})$ is below the heterotic Hagedorn temperature, \ie $R_{{\rm h}0} > R_{{\rm h}{\rm H}} \equiv (1+\sqrt{2})/\sqrt{2}$.  As shown in the appendix, the heterotic partition function can be brought into a form divided in three parts as follows :
\be
\label{Zh2}
\begin{array}{ll}
\dis Z_{\rm h}= {\h \beta_{\rm h} \hat V_{\rm h}\over \h\beta_{\rm h}^D} \Bigg\{ &\!\!\!\! \dis s_0b_0\, c_D+ \sum_{i=D}^9 2s_0b_{-1}\, G\bigg(2\pi R_{{\rm h}0}\Big\abs {1\over R_{{\rm h}i}}-R_{{\rm h}i}\Big\abs\bigg) \\
&\dis \!\!\!  \!+\!\sum_{\scriptsize \substack{A\ge 0, \, \b A\ge -1, \,  \v m, \, \v n\\ A-\b A=\v m\cdot \v n\\ (A,\v m,\v n)\neq (0,\epsilon \v e_i, \epsilon \v e_i),\\ \forall i, \forall \epsilon=-1,0,1}} \!\!\!\!s_Ab_{\b A}\, G\bigg(2\pi R_{{\rm h}0}\Big[4A+\sum_{j=D}^9\Big({m_j\over R_{{\rm h}j}}-n^jR_{{\rm h}j}\Big)^2\Big]^{1\over 2}\bigg)\Bigg\},
\end{array}
\ee
where the degeneracy $b_{\b A}$ of the right-moving bosonic string oscillator modes is defined from level $-1$.  The first contribution in $Z_{\rm h}$ is associated to the massless states labeled by $(A,\v m, \v n)=(0,\v 0,\v 0)$. They arise from the $\N_{10}=1$ supergravity and $SO(32)$ super-vector multiplets in ten dimensions. The second contribution comes from modes whose masses can vanish at particular values of the internal radii. For each $i$, these states are labeled as $(A,\v m, \v n)=(0, \epsilon\,  \v e_i,\epsilon \, \v e_i)$, where $\epsilon=\pm 1$ and $\v e_i$ is the unit vector in the direction $i$. They are massless at the self-dual point $R_{{\rm h}i}=1$, where they enhance the gauge symmetry, $U(1)\to SU(2)$.  The last line in (\ref{Zh2}) arises from the states which are never massless. It becomes substantial when Kaluza-Klein (winding) states become light, in the regime where  some $R_{{\rm h}i}$'s are large (small) compared to $2\pi R_{{\rm h}0}$ ($1/(2\pi R_{{\rm h}0})$). All other modes, being always super heavy as compared to the temperature scale, yield to exponentially suppressed contributions (see Eq.  (\ref{G})).


\vspace{-.3cm}
\subsection*{\it Duality map}
\vspace{-.3cm}
In ten dimensions, the heterotic/type I S-duality identifies the Einstein frame metrics and inverts the string couplings, $\lambda_{\rm h}=e^{\phi_{\rm h}^{(10)}}=e^{-\phi_{\rm I}^{(10)}}=1/\lambda_{\rm I}$ \cite{Polchinski:1995df}. In lower dimension $D$, these relations translate into a dictionary between the Einstein frame metrics, the internal radii and dilatons \cite{allD} :
\be
\label{hI}
\begin{array}{l}
ds^2_{{\rm h}(D)}=ds^2_{{\rm I}(D)}\\
\dis R_{{\rm h}i}={R_{{\rm I}i}\over \sqrt{\lambda_{\rm I}}}\equiv R_{{\rm I}i}\, {e^{-{1\over 2}\phi_{\rm I}^{(D)}}\over \left(\prod_{j=D}^92\pi R_{{\rm I}j}\right)^{1/4}}\; ,\quad i=0 \mbox{ or } D,...,9,\\
\dis \phi_{\rm h}^{(D)}=-{D-6\over 4}\, \phi_{\rm I}^{(D)}-{D-2\over 8}\sum_{i=D}^9\ln\left(2\pi R_{{\rm I}i}\right),
\end{array}
\ee
where the $D$-dimensional dilatons are defined as $\phi_{{\rm h},{\rm I}}^{(D)} =\phi_{{\rm h},{\rm I}}^{(10)}-{1\over 2}\sum_{i=D}^9 \ln(2\pi R_{{\rm h},{\rm I}i})$. Note that the Euclidean radii $R_{{\rm I}0}$ and $R_{{\rm h}0}$ are included in the above relations.  The inverse maps, which relate the type I fields in terms of heterotic quantities, are obtained by exchanging the subscripts ${\rm h}\leftrightarrow {\rm I}$.

We consider non-trivial evolutions for the Einstein frame metric, dilaton and internal radii moduli.  It is easily checked that the tree level heterotic and type I actions match, under the S-duality transformation (\ref{hI}) (\ie $S_{\rm h}^{\rm tree}=S_{\rm I}^{\rm tree}$).  The one-loop finite temperature effective potentials were computed using Euclidean backgrounds with laps functions $R_{{\rm h},{\rm I}0}$ in the string frames.  For the Lorentzian Einstein frame metric,
\be
\label{metric}
\begin{array}{l}
\dis ds_{(D)}^2={1\over (2\pi)^2}\left[ -\beta(x^0)^2{dx^0}^2+a(x^0)^2\left({dx^1}^2+\cdots +{dx^{D-1}}^2\right)\right]\\
\dis \beta = 2\pi R_{{\rm h},{\rm I}0}\, e^{-{2\over D-2}\phi_{h,{\rm I}}^{(D)}}\; , \qquad a^{D-1} = \h V_{h,{\rm I}}\, e^{-{2(D-1)\over D-2}\phi_{h,{\rm I}}^{(D)}} \, ,
\end{array}
\ee
the corresponding first order correction to the tree level action $S_{{\rm h},{\rm I}}^{\rm tree}$ is given by
\be
S_{{\rm h},{\rm I}}^{\mbox{\scriptsize 1-loop}}= -\int d^Dx \sqrt{-g^{(D)}}\,  \F_{{\rm h},{\rm I}} \qquad \where\qquad  \F_{{\rm h},{\rm I}}=-{Z_{{\rm h},{\rm I}}\over \beta a^{D-1}}.
\ee
Note that we do not  distinguish between the heterotic and type I inverse temperature $\beta$ and scale factor $a$ in (\ref{metric}), as they are measured in Einstein frame and are identified under the duality map (\ref{hI}). To be exactly equivalent, the effective actions should be corrected to all orders in perturbation theory and include non-perturbative effects as well. In the following, we will consider the heterotic point of view at weak coupling, $e^{\phi_{\rm h}^{(D)}}\!\ll 1$, restrict our computations at the one-loop order, and deduce the type I behavior in the dual regime.


\vspace{-.3cm}
\subsection*{\it Dual type I cosmological evolution}
\vspace{-.3cm}
To start, we apply the duality map to (\ref{Zh2}) and note that the first term exactly matches the sum of the first contributions in (\ref{ZT}) and (\ref{ZA}). This follows from the equality $b_0 = s_0 + (N^2 - N)/2$ and is due to the fact that the supergravity and $SO(32)$ super-vector multiplets are short, with protected vanishing masses.
Next, we concentrate on the interpretation and cosmological implications of the remaining terms in (\ref{Zh2}).
In the weakly coupled heterotic string, $e^{\phi_{\rm h}^{(D)}}\!\ll 1$, the time evolution of the universe for arbitrary initial conditions at the exit of the Hagedorn era can be analyzed along the lines of Refs  \cite{cosmoD,cosmoDb}. We first summarize the results here and then derive the dual type I cosmological behavior :

$\bullet$ When all radii satisfy $\abs R_{{\rm h}i}-1/R_{{\rm h}i} \abs < 1/(2\pi R_{{\rm h}0})$, $i=D,...,9$, the heterotic free energy density derived from (\ref{Zh2}) takes the form :
\be
\label{Zhe1}
\F_{\rm h}= -T^D \left\{ s_0b_0\, c_D+ \sum_{i=D}^9 2s_0b_{-1}\, G\bigg(2\pi R_{{\rm h}0}\Big\abs {1\over R_{{\rm h}i}}-R_{{\rm h}i}\Big\abs\bigg)+\O(e^{-2\pi R_{{\rm h}0}})\right\}.
\ee
Thanks to the properties (\ref{G}), the states with quantum numbers $(A,\v m,\v n)=(0,\epsilon \v e_i,\epsilon \v e_i)$ are responsible for the existence of a local minimum of $\F_{\rm h}$ at $R_{{\rm h}D}=\cdots=R_{{\rm h}9}=1$. The internal radii can be attracted and stabilized at this $SU(2)^{10-D}$ enhanced symmetry point. Moreover, for $D\ge 5$ the string coupling $e^{\phi_{\rm h}^{(D)}}$ (and thus $\lambda_{\rm h}$) freezes to some constant value $e^{\phi^{(D)}_{{\rm h}0}}$ determined by the initial conditions. For $D=4$, the dilaton $\phi_{\rm h}^{(4)}$ does not converge to a constant but instead decreases logarithmically with cosmological time. We show this in section \ref{stab} in a general context where we take into account all internal moduli. The rest of this section is valid for $D\ge 5$, while for $D=4$ one has to keep in mind the late time evolution of $\phi_{\rm h}^{(4)}$.

Applying the duality map (\ref{hI}), the ratios of the masses of the above winding-momentum states to the temperature become :
\be
\frac{\h M_{{\rm h}i}}{\h T_{\rm h}} \equiv 2\pi R_{{\rm h}0}\Big\abs {R_{{\rm h}i}-{1\over R_{{\rm h}i}}}\Big\abs = 2\pi R_{{\rm I}0}\Big\abs {{R_{{\rm I}i}\over \lambda_{\rm I}}-{1\over R_{{\rm I}i}}}\Big\abs \equiv {\h M_{{\rm I}i} \over \h T_{{\rm I}i}}\, .
\ee
From the type I point of view, the corresponding  BPS states have a natural interpretation as D (or anti-D)-strings wrapped once along the circles $S^1(R_{{\rm I}i})$, with one unit of momentum.
The heterotic cosmology translates into the type I context as follows. Whenever the type I radii start out in the dual range $\abs R_{{\rm I}i}/\lambda_{\rm I} -1/R_{{\rm I}i} \abs < 1/(2\pi R_{{\rm I}0})$, the light D-string modes can stabilize them at the point
\be
R_{{\rm I}i}=\sqrt{\lambda_{{\rm I}0}}\; , \qquad i=D,\dots, 9,
\ee
where $\lambda_{{\rm I}0}=1/\lambda_{{\rm h}0}\gg 1$ is the late time constant value of the  string coupling in ten dimensions. This implies the open string cosmology is well understood in type I, rather than in the T-dual picture in type I'. At each time, the width of the symmetric well of the potential for $\ln R_{{\rm I}i}$ is $\sqrt{\lambda_{\rm I}}/(2\pi R_{{\rm I}0})$ (see figure  \ref{pot}). In total, if we denote by $\phi_{{\rm I}0}^{(D)}$ the asymptotic value of the type I dilaton  in $D$ dimensions and use the inverse relations (\ref{hI}), the moduli are found to converge as follows,
\be
\label{cosmoI}
e^{\phi_{\rm I}^{(D)}(t)}\longrightarrow e^{\phi_{{\rm I}0}^{(D)}}\equiv  {e^{-{D-6\over 4}\phi_{{\rm h}0}^{(D)}}\over (2\pi)^{(10-D)(D-2)\over 8}}\; , ~~~~ R_{{\rm I}i}(t)\longrightarrow e^{{2\over D-6}\phi_{{\rm I}0}^{(D)}}\, (2\pi)^{10-D\over D-6}={1\over e^{{1\over 2}\phi_{{\rm h}0}^{(D)}}\, (2\pi)^{10-D\over 4}},
\ee
while the temperature and scale factor asymptotic behaviors are those of a radiation dominated era, $T^{-1}(t) \sim a(t)\sim t^{2/D}$, where $t$ is the cosmological time. Some remarks are in order :
\renewcommand{\labelitemi}{$\diamond$}
\begin{itemize}
\item For $D>6$, (\ref{cosmoI}) shows that the type I cosmology is at strong coupling. In this regime, solitons are generically light and the need to include their effects in the low energy effective action is not surprising.
\item For $D=6$, the asymptotic values of the moduli are $e^{\phi_{{\rm I}0}^{(6)}}=1/(2\pi)^2$ and $R_{{\rm I}i}(t)\to e^{-{1\over 2}\phi_{{\rm h}0}^{(6)}}/(2\pi)$. The type I picture is perturbative.
\item For $D<6$, the type I cosmological evolution is at weak coupling.  However, we observe the necessity to take into account the contributions arising from solitons which are light, when we sit in the neighborhood of the enhanced symmetry points.
\end{itemize}
\renewcommand{\labelitemi}{$\bullet$}
In summary, for $D\neq 6$ in type I, the internal radii are stabilized while the dilaton $\phi_{\rm I}^{(D)}$ freezes somewhere along its flat direction. On the contrary, for $D=6$, the dilaton is stabilized, all complex structures $R_{{\rm I}i}/R_{{\rm I}j}$ are also stabilized at one, while the internal space volume $\prod_{i=D}^9(2\pi R_{{\rm I}i})$ freezes along a flat direction.  This is not a surprise, since in $D=6$ the heterotic/type I duality exchanges internal volumes and string couplings : $\prod_{i=D}^9(2\pi R_{{\rm h},{\rm I}i})\leftrightarrow 1/e^{2\phi_{{\rm I},{\rm h}}^{(6)}}$.

$\bullet$ If at some epoch one of the  heterotic internal radii satisfies  $R_{{\rm h}j}>2\pi R_{{\rm h}0}$, while the $9-D$ remaining ones are stabilized, $R_{{\rm h}i}= 1$ for $i\neq j$, the free energy density deduced from (\ref{Zh2}) becomes
\be
\label{>}
\F_{\rm h}=-T^D\big( s_0b_0+(9-D)\, 2s_0b_{-1}\big)\bigg[c_D+\sum_{m_{j}\neq 0}G\Big(2\pi R_{{\rm h}0}{\abs m_{j}\abs \over R_{{\rm h}j}}\Big)\bigg]+\O(e^{-2\pi R_{{\rm h}0}}).
\ee
We see that in addition to the massless supergravity and $SO(32)$ super-vector multiplets, there are also contributions coming from their Kaluza-Klein descendants,  which are light since $R_{{\rm h}j}$ is large.  Applying the duality rules and comparing to the perturbative type I result in the first line of (\ref{FI}), we observe a match up to an additional contribution $(9-D)\, 2s_0b_{-1}$ to the overall numerical coefficient.  This discrepancy arises from the extra massless D (or anti-D)-strings responsible for the stabilization of the $R_{{\rm I}i}$'s at $\sqrt{\lambda_{\rm I}}$.
Therefore, the main difference with the pure perturbative analysis is that the plateau of the effective potential is lowered and that the slope for $R_{{\rm I}j}>2\pi R_{{\rm I}0}$ is steeper (see figure \ref{pot}). The cosmological evolution is however similar to the one discussed below (\ref{range}).  As their heterotic counterparts \cite{cosmoD,cosmoDb}, $R_{{\rm I}j}(t)$ and $R_{{\rm I}0}(t)$ evolve such that the regime where $R_{{\rm I}j}(t)<2\pi R_{{\rm I}0}(t)$ is reached. After that, $R_{{\rm I}j}$ freezes along its plateau or is stabilized at $\sqrt{\lambda_{\rm I}}$ as explained before.

$\bullet$ In a similar way, if a heterotic radius satisfies $R_{{\rm h}j}<1/(2\pi R_{{\rm h}0})$, while the others are stabilized at their self-dual points, $R_{{\rm h}i}=1$ for $i\neq j$,  we have
\be
\label{<}
\F_{\rm h}=-T^D\big( s_0b_0+(9-D)\, 2s_0b_{-1}\big)\bigg[c_D+\sum_{n_{j}\neq 0}G\Big(2\pi R_{{\rm h}0}\abs n_{j}\abs R_{{\rm h}j}\Big)\bigg]+\O(e^{-2\pi R_{{\rm h}0}}).
\ee
In this case, substantial  contributions arise from the winding modes along $S^1(R_{{\rm h}j})$, which are light since $R_{{\rm h}j}$ is small enough.
Their effect is to attract $R_{{\rm h}j}(t)$ to values larger than $1/(2\pi R_{{\rm h}0}(t))$ \cite{cosmoD,cosmoDb}.  Applying the S-duality rules to translate this statement in the type I context, we find that
if $R_{{\rm I}j}< \lambda_{\rm I}/(2\pi R_{{\rm h}0})$ at some time, the evolution of these moduli implies we end in a regime where $\lambda_{\rm I}/(2\pi R_{{\rm h}0})<R_{{\rm I}j}$, after which the internal modulus freezes or is stabilized at $\sqrt{\lambda_{\rm I}}$. Noting that the argument of the $G$-function in (\ref{<}) becomes
\be
{\h M_{{\rm h}j}\over \h T_{\rm h}} \equiv 2\pi R_{{\rm h}0}\, \abs n^j\abs R_{{\rm h}j}=2\pi R_{{\rm I}0}\, \abs n^j\abs {R_{{\rm I}j}\over \lambda_{\rm I}}\equiv {\h M_{{\rm I}j}\over \h T_{\rm I}},
\ee
we conclude that  the above mechanism is due to two sets of towers of D-string winding modes along $S^1(R_{{\rm I}j})$. The first one contains ``solitonic descendants" of the perturbative massless supergravity and $SO(32)$ super-vector multiplets.
The second set is associated to the descendants of the D (or anti-D)-strings responsible for the stabilization of the $(9-D)$ internal radii $R_{{\rm I}i}$ at $\sqrt{\lambda_{\rm I}}$.
The net result of these non-perturbative light states is to render the type I free energy explicitly invariant under the ``non-perturbative T-duality" $R_{{\rm I}j}\to \lambda_{\rm I}/R_{{\rm I}j}$ (see figure \ref{pot})\footnote{Since at late times $\lambda_{\rm I}(t)\to \lambda_{{\rm I}0}$ and $R_{{\rm I}0}(t)\to +\infty$, the left-boundary of the plateau of the effective potential of $\ln R_{{\rm I}j}$ ends by being negative. This means that $R_{{\rm I}j}$ may freeze at some value below one. In such a case, a T-duality $R_{{\rm I}j}\to 1/R_{{\rm I}j}$ to a type I' description is more suitable. In general, a mixed type I / type I' theory may  be obtained, in order to keep all internal radii larger than one.}.


\vspace{-.3cm}
\subsection*{\it Comments}
\vspace{-.3cm}
To conclude this  section, we would like to make some remarks. We first observe that under the duality map (\ref{hI}), the Hagedorn radii do not match. We thus infer from the perturbative heterotic side a new value of the Hagedorn radius in type I, when $\lambda_{\rm I}$ is large :
\be
R_{{\rm I}{\rm H}} = \left\{ \begin{array}{cl}
        \sqrt{2}&\;\;\;\;\mbox{ for } \lambda_{\rm I} \ll 1 \\
        \sqrt{\lambda_{\rm I}} \, \frac{1 + \sqrt{2}}{\sqrt{2}}&\;\;\;\;\mbox{ for } \lambda_{\rm I} \gg 1
      \end{array} \right.\cdot
\ee
From a cosmological point of view, $R_{{\rm I}{\rm H}}$ in the regime $\lambda_{\rm I}(t) \gg 1$ is thus a time-dependent scale. Note that this non-perturbative expression for $R_{{\rm I}{\rm H}}$ obtained once D-strings are taken into account can be relevant even at weak coupling, $e^{\phi_{\rm I}^{(D)}}\!\ll 1$. This is for instance the case  for $D\le 6$, when $\sqrt{\lambda_{\rm I}}$ and the $R_{{\rm I}i}$'s reach the asymptotic value $\sqrt{\lambda_{{\rm I}0}}\gg 1$.

For $D\ge 7 $, the stabilization of the internal type I radii at $\sqrt{\lambda_{{\rm I}0}}\gg 1$ occurs at strong coupling,  $e^{\phi_{\rm I}^{(D)}}\!\gg 1$. However, the D-string states responsible for this effect are BPS, so that their masses are protected by supersymmetry. Thus, these modes remain massless for arbitrary $\lambda_{\rm I}$, when $R_{{\rm I}i}=\sqrt{\lambda_{{\rm I}}}$. It follows that the type I free energy density can easily be determined when $\lambda_{\rm I} \ll 1$ and $R_{{\rm I}i}\simeq \sqrt{\lambda_{{\rm I}}}$.  It is actually given by (\ref{Zhe1}), once translated in terms of dual type I variables. The justification of this statement is based on the following facts. In this regime, the string coupling is weak,  $e^{\phi_{\rm I}^{(D)}}\!\ll 1$, and the contribution of the perturbative part of the spectrum is that of a perfect gas. Moreover, the contribution of the light solitons is of identical form, since $SU(2)$'s (gauge) symmetries transform them into the perturbative modes in the Cartan subalgebras. We conclude that the mechanism of stabilization of the internal type I radii remains valid at weak coupling  $e^{\phi_{\rm I}^{(D)}}\!\ll 1$. Since this yields $R_{{\rm I}i}=\sqrt{\lambda_{{\rm I}0}}\ll 1$, the model is better described in the T-dual type I' picture. However, the dynamics in the intermediate regime $e^{\phi_{\rm I}^{(D)}}\simeq 1$ for $D\ge 7$ (or $e^{\phi_{\rm I}^{(D)}}\!\ll\!\!\! \!\!\! / \;\,1$  for $D\le 6$) cannot be inferred from these arguments.

Finally, for $D \le 5 $, additional non-perturbative states may play a role in the cosmological evolution.  In fact, D5-branes of the type I theory (or NS5-branes in the heterotic context) can wrap the internal manifold in analogy with the D-strings we have considered\footnote{Note that these states may contribute even for $D=5$.  This is to be contrasted with 5-brane instantons at zero temperature, which require an internal space of six dimensions.}. It would be interesting to study their effects on the dynamics, which may lead eventually to a  stabilization of the dilaton.


\section{E1-instanton corrections}
\label{E1}

We have found that non-perturbative states contribute to the type I free energy density. In the literature, corrections to the low energy effective action are often considered from another point of view, namely instantons and their stringy generalizations. For instance, E1 contributions to holomorphic couplings have been analyzed in supersymmetric cases by heterotic/type I duality \cite{Bachas:1997mc}. In the present section, our aim is to reexamine the type I free energy from the point of view of E1-instantons and single out the configurations responsible for the stabilization of internal radii. In this non-supersymmetric case, we want to predict the E1 corrections in type I from dual heterotic worldsheet instantons. For simplicity, we restrict our analysis to the case $D=9$, where instantons wrap the Euclidean time circle and the direction 9. This is to be contrasted with the zero temperature case where E1 corrections would only arise for $D\le 8$.  We note that by a double Wick rotation, the results in this section may be interpreted as the zero temperature vacuum energy contribution of E1-instantons wrapping an internal $T^2$, with spontaneous supersymmetry breaking boundary conditions along one of the toroidal directions.  In this case, the temperature scale $T$ is replaced with the supersymmetry breaking scale $M$.

Our starting point is the heterotic model of section \ref{dual}. To help exhibit the worldsheet instanton structure of the one-loop amplitude $Z_{\rm h}$, we work in the Lagrangian formulation of the zero modes lattice associated to $S^1(R_{{\rm h}0}) \times S^1(R_{{\rm h}9})$ (see Eqs (\ref{ZhApp1}) and (\ref{poisson})). We consider $R_{{\rm h}9}\ge 1$ and parameterize the zero modes by the matrix $\M=\Big(\!\!\begin{array}{lr}n^0&\t m_0\\ n^9&\t m_9\end{array}\!\!\Big)$. The case $R_{{\rm h}9} \leq 1$ may be obtained by T-duality.    We may decompose the lattice sum under orbits of the  $SL(2,\Z)$ modular group   as follows.  For any set of  modular covariant functions $f_{\M}(\tau,\bar\tau)$ such that $f_\M(M(\tau),M(\b \tau))=f_{\M M}(\tau,\b \tau)$, for all $M\in SL(2,\Z)$, one has
\be
\label{unfo2}
\begin{array}{ll}
\dis \int_\F{d^2\tau\over \tau_2^2}\, \sum_{\M}f_\M(\tau,\b \tau)=&\!\!\!\dis \int_\F{d^2\tau\over \tau_2^2}\, f_{\big(\!\!\!\!\tiny\begin{array}{l}0\,0\\0\,0\end{array}\!\!\!\!\big)}(\tau,\b \tau)\\
&\!\!\!\dis +\int_{\S_+}{d^2\tau\over \tau_2^2}\, {\sum_{\t m_0,\t m_9}}^{\!\!\prime}f_{\big(\!\!\!\!\tiny\begin{array}{l}0\,\t m_0\\0\,\t m_9\end{array}\!\!\!\!\big)}(\tau,\b \tau)+\int_{\Co_+}{d^2\tau\over \tau_2^2}\,  2\!\!\!\!\!\sum_{\small \substack{\t m_0\neq 0 \\ n^9>\t m_9\ge 0}}\!\!\!f_{\big(\!\!\!\!\tiny\begin{array}{l}\,0\;\;\t m_0\\n^9\,\t m_9\end{array}\!\!\!\!\big)}(\tau,\b \tau).
\end{array}
\ee
This is easily shown by applying Eq. (\ref{unfo}) twice : First to the sum over $(n^0,\t m_0)$ and then to the sum over $(n^9,\t m_9)$. The integral over the upper half plane $\Co_+$ is obtained for $n^9>0$ by writing $\t m_9=kn^9+l$ ($0\le l<n^9-1$) and changing $\tau\to\tau+k$. The integral over $\F$ corresponds to the zero orbit (\ie $\M = 0$), while the integral over $\S_+$ corresponds to non-vanishing degenerate matrices (\ie with $\det \M=0$).  The last integral over $\Co_+$ is associated to non-degenerate matrices.

Applying (\ref{unfo2}) to the heterotic partition function $Z_{\rm h}$, the contribution of the zero orbit vanishes due to supersymmetry, so that\footnote{
The use of Eq. (\ref{unfo2}) is valid if the argument of the discrete sum to integrate is absolutely convergent. In the present case, since the right-moving block $\Gamma_{(0,16)}/\b\eta^{24}$ and the left-moving $O_8/\eta^8$ character involve diverging powers of $e^{2\pi\tau_2}$ in the limit $\tau_2\to +\infty$, Eq. (\ref{unfo2}) can be trusted if $R_{{\rm h}0}>\sqrt{3}$ and $R_{{\rm h}9}>\sqrt{2}$.  The first condition is not problematic as we are focussing on the dynamics at low temperature. Since we are interested in the stabilization of $R_{{\rm h}9}$ around $1$, the second condition could be a problem. However, we see shortly that the final expression (\ref{Znd}) can be analytically continued all the way to $R_{{\rm h}9} = 1$.}
\begin{eqnarray}
\nonumber &&Z_{\rm h}= Z_{\rm h}^d+Z_{\rm h}^{nd}\\
&&\label{Zdnd} Z_{\rm h}^d= {\h \beta_{\rm h} \hat V_{\rm h}\over (2\pi)^9}\int_{\S_+}{d^2\tau\over 2 \tau_2^6}\, {\Gamma_{(0,16)}\over \eta^8\b\eta^{24}}\, R_9{\sum_{\t m_0,\t m_9}}^{\!\!\prime}\,\,e^{-{\pi R_{{\rm h}0}\over \tau_2}\t m_0^2}\, e^{-{\pi R_{{\rm h}9}\over \tau_2}\t m_9^2}\Big[V_8-(-1)^{\t m_0}S_8\Big]\\
&&\nonumber Z_{\rm h}^{nd}= {\h \beta_{\rm h} \hat V_{\rm h}\over (2\pi)^9}\int_{\Co_+}{d^2\tau\over 2 \tau_2^6}\, {\Gamma_{(0,16)}\over \eta^8\b\eta^{24}}\, R_9\, 2\!\!\!\!\!\sum_{\small \substack{\t m_0\neq 0 \\ n^9>\t m_9\ge 0}}\!\!\!\!e^{-{\pi R_{{\rm h}0}\over \tau_2}\t m_0^2}\, e^{-{\pi R_{{\rm h}9}\over \tau_2}\abs n^9\tau+\t m_9\abs^2} \Big[V_8-(-1)^{\t m_0}S_8\Big].
\end{eqnarray}
Performing the $\tau$-integrations, the degenerate part $Z_{\rm h}^{d}$ can be brought into the form
\be
\label{Zd}
Z_{\rm h}^d={\h \beta_{\rm h} \hat V_{\rm h}\over \h \beta_{\rm h}^9} \Bigg\{  s_0b_0\, c_9+\, {\sum_{A\ge 0,\, m_9}}^{\!\!\!\!\prime}\;\;s_Ab_A\, G\bigg(2\pi R_{{\rm h}0}\Big[4A+\Big({m_9\over R_{{\rm h}9}}\Big)^2 \Big]^{1\over 2}\bigg)\Bigg\} \, ,
\ee
while the non-degenerate contribution $Z_{\rm h}^{nd}$ can be written as,
\be
\label{Znd}
Z_{\rm h}^{nd}= {\h \beta_{\rm h} \hat V_{\rm h}\over \h \beta_{\rm h}^9}\;\;\; 2\!\!\!\!\!\!\sum_{\small \substack{A\ge 0,\, \b A \ge -1\\ n^9>\t m_9\ge 0}}\!\!s_Ab_{\b A}\, {e^{-2i\pi{\t m_9\over n^9}(A-\b A)}\over n^9}\, G\bigg(2\pi R_{{\rm h}0}\Big[4A+\Big({A-\b A\over n^9R_{{\rm h}9}}-n^9R_{{\rm h}9}\Big)^2\Big]^{1\over 2}\bigg).
\ee
Summing over $\tilde m_9$ in (\ref{Znd}) enforces the level matching condition $A-\b A = n^9 m_9$ for some integer $m_9$, whenever $n^9\neq 0$. The ``missing term'' for $n^9=0$ is actually the contribution of the degenerate orbits $Z_{\rm h}^{d}$.  In total, $Z_{\rm h}^d+Z_{\rm h}^{nd}$ yields with no surprise the expression  (\ref{Zh2}), which can be analytically continued in  the range $1\le R_{{\rm h}9}\le \sqrt{2}$.   However, to exhibit the instantonic structure, it is better to leave the sum over $\tilde m_9$.

In $Z_{\rm h}^d$, only pure Kaluza-Klein modes along the directions 9 and 0 contribute and the worldsheet embedding in the target torus is trivial (no instanton number).  Therefore, these states do not play a role in stabilizing the internal circle.  In order to extract the configurations in $Z_{\rm h}^{nd}$ responsible for fixing $R_{{\rm h}9}$ at the self-dual point, we know it is enough to focus on the dominant contributions in the low temperature expansion.  The terms with $A\ge 1$ are exponentially suppressed, $\O(e^{-4\pi R_{{\rm h}0}})$, compared to the contribution with $A=0$. The latter arises from BPS configurations and, at this level of approximation, $Z_{\rm h}^{nd}$ in Eq. (\ref{Zdnd}) involves a purely antiholomorphic function, $\B(\b \tau) =\Gamma_{(0,16)}/\b\eta^{24}$, dressed by an inverse power of $\tau_2$ and the lattice of zero modes associated to the directions 0 and 9.  This form is similar to the one encountered in the evaluation of holomorphic couplings, when supersymmetry is unbroken \cite{Bachas:1997mc}.

We can now define instanton configurations, with associated \Ka and complex structure moduli $\Upsilon$ and $\Y$ as,
\be
\mbox{\em Instanton with } ~n^9> \t m_9\ge 0\; , \; \,\t k_0\ge 0 : ~\left\{\begin{array}{l}\Upsilon=i\Upsilon_2=i(2\t k_0+1)R_{h0}\cdot n^9R_{h9}\\ \dis\Y=\Y_1+i\Y_2={\t m_9\over n^9}+i\, {(2\t k_0+1)R_{h0}\over n^9R_{h9}}
\end{array}\right. ,
\ee
where $(2\t k_0+1)\, n^9$ is the instanton number, which counts the number of times the worldsheet wraps around the target torus. Using these notations and introducing coefficients $\alpha_n\in \mathbb{N}$ in the expansion of the Bessel function\footnote{In any odd dimension, the Bessel function admits a power series with a finite number of terms.} in (\ref{cg}), $K_{9\over 2}(x)=\sqrt{\pi/(2x)} e^{-x}\sum_{n=0}^4 \alpha_n/x^n$,
we may write (\ref{Znd}) as
\be
\label{Zndins}
Z_{\rm h}^{nd}={\h V_{\rm h}^{(10)}\over  (2\pi)^{10}} \; 2\!\!\!\!\!\sum_{\rm \scriptsize instantons}\!\!\!\!s_0\, {e^{2i\pi\Upsilon}\over \Upsilon_2\, \Y_2^4}\, \sum_{n=0}^4\left[{\alpha_n\over (2\pi \Upsilon_2)^n} \! \sum_{\b A\ge -1}b_{\b A}\left(1+\b A\, {\Y_2\over \Upsilon_2}\right)^{4-n}\!\!e^{2i\pi \Y\b A}\right]\!+c.c. +\O(e^{-4\pi R_{h0}}),
\ee
where $\h V_{\rm h}^{(10)}$ is the ten-dimensional Euclidean volume. This result can be given a more elegant appearance by noting that $\B(\Y)$ is a modular form of weight 4. Introducing the modular covariant derivative $D\X=(\partial_\Y+{ir\over 2\Y_2})\X$, where $\X(\Y)$ is any modular form of weight $r$,\footnote{This means that $\X(\Y+1)=\X(\Y)$ and $\X(-1/\Y)=\X(\Y)/\Y^r$. Moreover, $D\X$ is a modular form of weight $r-2$.} the brackets in (\ref{Zndins}) become ${1/(\pi \Upsilon_2)^n}\sum_{m=0}^n\gamma_{nm}(i\Y_2)^mD^m\B(\Y)$, where $\gamma_{nm}$ are rational numbers.

The above expression of $Z_{\rm h}^{nd}$ contains far too many explicit terms needed to study the stabilization of $R_{{\rm h}9}$. In ({\ref{Znd}), the dominant contribution for $A=0$ arises when $\b A=-1$ and $n^9=1$, while the remaining terms are exponentially suppressed, $\O(e^{-2\pi R_{{\rm h}0}})$.  Restricting to $\b A=-1$ and the instanton configurations $n^9=1$, $\t m_9=0$, $\t k_0\ge 0$ in $Z^{nd}_{\rm h}$, we can add the degenerate contribution $Z_{\rm h}^d=(\h \beta_{\rm h} \hat V_{\rm h}/\h \beta_{\rm h}^9)\, s_0b_0\, c_9+\O(e^{-2\pi R_{{\rm h}0}})$ to recover the first line of Eq. (\ref{Zh2}) required for the derivation of the stabilization of $R_{{\rm h}9}$.

We now wish to interpret Eq. (\ref{Zndins})  from the perspective of the type I superstring.  Under the heterotic/type I dictionary (\ref{hI}), the complex and \Ka structures  $\Y$ and $\Upsilon$ are mapped into $\Y_{\rm I}$ and $\Upsilon_{{\rm I}}/\lambda_{\rm I}$. Consequently, the exponential factor of $\Upsilon$ in (\ref{Zndins}) yields the exponential of the Nambu-Goto action for a D-string, and $Z_{\rm h}^{nd}$ translates into a sum of E1 instantons as in \cite{Bachas:1997mc},
\be
\label{ZndinsI}
Z_{\rm I}^{E1}\!=\!{\h V_{\rm I}^{(10)}\over  (2\pi)^{10}} \; 2\!\!\!\!\!\!\!\!\sum_{\rm \scriptsize E1 \, instantons}\!\!\!\!\!\!\!\!s_0\, {e^{\frac{2i\pi}{\lambda_{\rm I}}\Upsilon_{\rm I}}\over \Upsilon_{{\rm I}2}\, \Y_{{\rm I}2}^4} \sum_{n=0}^4\!\!\left[\!{\alpha_n\over (2\pi \Upsilon_{{\rm I}2})^n} \!\! \sum_{\b A\ge -1}\!\!b_{\b A}\!\left(\!\frac{1}{\lambda_{\rm I}}+\b A\, {\Y_{{\rm I}2}\over \Upsilon_{{\rm I}2}}\!\right)^{4-n}\!\!\!\!e^{2i\pi \Y_{\rm I}\b A}\!\right]\!+c.c. +\O(e^{-4\pi {R_{{\rm I}0}\over \sqrt{\lambda_{\rm I}}}}).
\ee
Actually, the configurations of the D-string worldsheets wrapped on $S^1(R_{{\rm I}0}) \times S^1(R_{{\rm I}9})$ are highly dissymmetric at late times in the sense that $R_{{\rm I}0}(t)\to +\infty$ and $R_{{\rm I}9}(t)\sim \sqrt{\lambda_{\rm I}(t)}\to \sqrt{\lambda_{{\rm I}0}}$.  However, this does not mean it is artificial to consider such E1-instantons.  Instead, they open the possibility to derive from a pure type I point of view the free energy responsible for the stabilization of the internal moduli (or the effective potential at zero temperature when at least two internal directions are compactified and supersymmetry is spontaneously broken). Thus, it would be interesting to derive D-brane instanton corrections from first principles, in the case where supersymmetry is spontaneously broken. The full instantonic structure of (\ref{Znd}) should also be interpreted from a type I point of view, even when all contributions with $A\ge 0$ and $\b A\ge -1$ are kept explicitly.


\section{Heterotic and dual type I moduli stabilization}
\label{stab}

We would like to extend the analysis used in section \ref{dual} to include the remaining moduli in addition to the internal radii.  We consider the heterotic string  compactified on $T^{10-D}$ at a generic point in moduli space and show that when finite temperature is switched on,  the free energy density can stabilize all internal moduli.
Our study is based on the effective action at finite temperature and weak coupling for the massless degrees of freedom, while all massive states are integrated out. Introducing simplified notations, we are interested in non-trivial backgrounds for the Einstein frame metric $g$, the dilaton $\phi$ in $D$ dimensions and all real-valued internal moduli $\Phi^M$, which we denote collectively as $\v \Phi$.  Concretely, $\v\Phi$ contains the components of the metric $\h g_{ij}$ and antisymmetric tensor $B_{ij}$, together with the Wilson lines $Y^I_{i}$ ($i,j=D,\dots,9$; $I=10,11,\dots,25$).  It is then straightforward to deduce the dynamics and final expectation values of the type I counterparts of these scalars by using the duality map
\be
\label{dico}
\h g_{ij}={\h g_{\rmI ij}\over \lambda_{\rmI}}\; , \qquad \qquad B_{ij}=C_{ij}\; , \qquad\qquad  Y_i^I=Y_{\rmI i}^I\; ,
\ee
where $C_{ij}$ is the RR 2-form.  Detailed examples of this analysis  will be given in section \ref{example} for  $D=8$.

The heterotic low energy effective action
\be
\label{stab81}
    S=\int d^Dx\sqrt{-g}\left[\frac{R}{2}-\frac{2}{D-2}\,  \partial_\mu \phi \partial^\mu \phi-\demi F_{MN} \partial_\mu \Phi^M \partial^\mu \Phi^N-\F \right]
\ee
involves the tree level moduli space metric $F_{MN}=F_{MN}(\v \Phi)$ and the one-loop free energy density $\F=\F(T,\phi,\v \Phi)$. Since the backreaction of $\F$ on the classical background is already a one-loop effect, there is no need to take into account the quantum corrections to the kinetic terms.
For homogeneous and isotropic evolutions, variation of $S$ with respect to the time-dependent metric (\ref{metric}), dilaton and moduli $\Phi^M$ yields, in cosmological time defined by $dt\equiv \beta(x^0) dx^0$,
\begin{align}
\label{stab9}
    &\frac{(D-1)(D-2)}{2}\, H^2=\frac{2}{D-2}\, \dot{\phi}^2+\demi\,  F_{MN}\dot{\Phi}^M\dot{\Phi}^N+\rho,\\
    \label{stab10}
    &\frac{(D-1)(D-2)}{2}\, H^2+(D-2)\dot{H}+\frac{2}{D-2}\, \dot{\phi}^2+\demi \, F_{MN}\dot{\Phi}^M\dot{\Phi}^N+P=0,\\ \label{stab11}
    &\ddot{\phi}+(D-1)H\dot{\phi} + \frac{D-2}{4}\, \cf_{\phi}=0,\\ \label{stab12}
    &\ddot \Phi^M+(D-1)H\dot\Phi^M+F^{MN}\Big(F_{NPQ}-{1\over 2}F_{PQN}\Big)\dot\Phi^P\dot\Phi^Q+F^{MN}\F_N=0.
\end{align}
In these equations, $H=\dot a/a$ and the thermal pressure and energy density are found to be
\be
P=-\F  \; , \qquad \qquad \rho=T\, \frac{\partial P}{\partial T}-P  .
\ee
Additional indices $\phi$ and $N$ denote partial derivatives with respect to $\phi$ and $\Phi^N$,  while $F^{MN}\equiv (F^{-1})^{MN}$.  It is convenient to replace Eq. (\ref{stab10}) by the constant entropy constraint. The  latter is found by integrating the energy-momentum tensor conservation law derived from the above differential system (see \cite{cosmoD}),
\be
\label{stab1310}
\dot{\rho}+(D-1)H(\rho+P)=\dot{\phi}\, \cf_{\phi}+\dot{\Phi}^M \cf_{M}\quad \Longrightarrow\quad a^{D-1}\,  {\rho+P\over T}=\text{constant entropy}.
\ee

In order to find particular evolutions characterized by static moduli, $\big(\phi(t),\v \Phi(t)\big)\equiv (\phi_0,\v \Phi_0)$, we need to specify $\F$.  For any supersymmetric spectrum, the one-loop free energy density is
\be
\label{freegene}
\F= - e^{{2D\over D-2}\phi}\!\int_0^{+\infty} {dl\over 2l}\, {1\over (2\pi l)^{D\over 2}}\,\sum_s e^{-{\h M_s^2 l\over2}}\, \sum_{\t m_0} e^{-\hat \beta^2\t m_0^2\over 2l}\left(1-(-)^{\t m_0}\right)= -T^D \sum_s G\bigg({e^{{2\over D-2}\phi}\h M_s\over T}\bigg),
\ee
where $\h M_s$ is the mass of each boson/fermion pair $s$, and the dilaton dressing in front of the integral is introduced to switch from string to Einstein frame.
This general expression applied to our case of interest, namely the heterotic string on $T^{10-D}$, is explicitly derived from a one-loop vacuum-to-vacuum amplitude in the appendix. In the notations introduced there, $s_0r_0=2^8\times 24$ boson/fermion pairs of states are massless everywhere in moduli space\footnote{They are associated to the supergravity and super-vector multiplets of the $SO(32)$ Cartan generators.}, while the other modes have moduli-dependent masses, $\h M_s(\v \Phi)$. Since light states have the tendency to lower $\F$, effective potential wells can be found at any point $\v \Phi_0$ where $n_0>0$ pairs of modes generically massive are accidentally massless, $\h M_u(\v\Phi_0)=0$, $u=1,\dots,n_0$. The fact that we have at zero temperature 16 real conserved supercharges implies that such points are associated to enhancements of the gauge symmetry.  Defining $\h M_{\rm min}$ to be the lightest non-vanishing mass at $\v \Phi_0$, the free energy density can be written in a neighborhood of $\v \Phi_0$ as,
\be
\F=-T^D\left\{ s_0r_0+\sum_{u=1}^{n_0}G\bigg({e^{{2\over D-2}\phi}\h M_u(\v\Phi)\over T}\bigg)+ \O\big(e^{-{\h M_{\rm min}\over \h T}}\big)\right\} .
\ee
At low enough temperature, the exponentially suppressed terms can be neglected and we may derive identities for the thermal source terms at $\v\Phi_0$, including the equation of state,
\be
\label{stab15}
\!\!\left.\cf\right|_{\v\Phi_0}=-T^D (s_0r_0+n_0)\, c_D\; , \quad \left.\cf_{\phi}\rst_{\v \Phi_0} =0\; , \quad \left. \cf_M\rst_{\v \Phi_0} = 0\; , \quad \rho|_{\v\Phi_0}=(D-1)P|_{\v\Phi_0}\propto T^D.
\ee
It is then straightforward to check that the evolutions
\be
\label{radera}
a_0(t) \propto {1\over T_0(t)} \propto t^{2/D}\; ,  \qquad\qquad \phi(t)\equiv \phi_0 \; , \qquad \qquad \v \Phi(t)\equiv \v\Phi_0\; ,
\ee
corresponding to radiation eras with static moduli are particular solutions of the equations of motion.

The above trajectories are actually attractors of the dynamics in some circumstances. To study this, we analyze their stability under small time-dependent deviations,
\be
\label{stab18}
    a=a_0(1+\epsilon_a)\; ,\qquad  T=T_0(1+\epsilon_T)\; ,\qquad  \phi=\phi_0+\epsilon_{\phi}\; ,\qquad  \Phi^M=\Phi^M_0+\epsilon^M .
\ee
We first perturb the internal moduli equation (\ref{stab12}). Denoting $H_0=\dot a_0/a_0$, one obtains at lowest order,
\be
\label{Ptbg10}
    \ddot{\epsilon}^M+(D-1)H_0\, \dot{\epsilon}^M +  {\Lambda^M}_{\!\!N}\, \epsilon^N=0\qquad \where\qquad {\Lambda^M}_{\!\!N}\equiv  F^{ML}\abs_{\v\Phi_0}\F_{LN}\abs_{(T_0,\phi_0,\v\Phi_0)}.
\ee
${\Lambda^M}_{\!\!N}$ is an effective ``time-dependant squared mass matrix" evaluated for the background (\ref{radera}). Since
\be
\label{2dder}
\F_{MN}\abs_{\v\Phi_0}=T^{D-2}\, e^{\frac{4\phi}{D-2}}\, \frac{c_{D-2}}{4 \pi} \sum_{u=1}^{n_0}\left.\frac{\partial^2\hat{M}_u^2}{\partial\Phi^M\partial\Phi^N}\right\abs_{\v\Phi_0}
\ee
is semi-definite positive, ${\Lambda^M}_{\!\!N}$ is diagonalizable with non-negative eigenvalues\footnote{This follows from the fact that the matrices $F^{-1/2}$ and $\F$ are (semi-)definite positive,  so that $F^{-1/2}\F F^{-1/2}=F^{1/2}\Lambda F^{-1/2}$ is semi-definite positive. Note that in models where the spontaneous breaking of supersymmetry is generic \ie not only due to thermal effects, each term in the sum over the boson-fermion pair $u$ in  Eq. (\ref{2dder}) is dressed with a $+$ (or $-$) sign when the boson (fermion) is lighter than the fermion (boson). In such cases, $\F$ is not semi-definite positive and the extrema of $\F$ can be minima, maxima or saddle points.}, which we define as ${4\lambda^2_M \over D^2t^{2(D-2)/D}}$. In the case when some $\lambda_M$'s vanish, one needs to take into account quadratic terms in Eq. (\ref{Ptbg10}) (see the discussion of the dilaton equation below). In particular, this is required when moduli sit on the plateau of their thermal effective potential (see figure  \ref{pot}). For simplicity, we proceed by analyzing the most interesting case, where all internal moduli are ``massive", which means $\lambda^M >0$. Switching to a diagonal basis of perturbations $\t \epsilon^M$, one  obtains from (\ref{Ptbg10})
\be
\label{Ptbg14}
 \tilde{\epsilon}^M = {t^{1/D}\over \sqrt{t}}\left[C^M_+\, J_{\frac{D-2}{4}}(\lambda_M\, t^{2/D}) +C^M_-\, J_{-\frac{D-2}{4}}(\lambda_M\, t^{2/D})\right],
\ee
where $C^M_\pm$ are integration constants and $J_{\pm {D-2\over 4}}$  are Bessel functions of the first kind\footnote{For $D=6$, $J_{-1}$ should be replaced by the Bessel function of the second kind, $Y_{-1}$.}. This describes damped oscillations with amplitude of order $1/\sqrt{t}$, where $t$ is supposed to be large enough so that $\abs\tilde {\epsilon}^M\abs \ll 1$  is satisfied.

Next, we derive from (\ref{stab11}) the equation for the dilaton perturbation at leading order,
\be
\label{dilper}
    (a_0^{D-1}\dot{\epsilon}_\phi)\!\dot{\phantom{\Phi}}\! +  a_0^{D-1}\, {1\over 2}\, \F_{\phi M N}\abs_{(T_0,\phi_0,\v\Phi_0)} \epsilon^M\epsilon^N=0\; \quad \where\; \quad \F_{\phi M N}\abs_{\v\Phi_0}\equiv  {4\over D-2}\, \F_{M N}\abs_{\v\Phi_0}.
\ee
Since the constants $C^M_\pm$ are a priori of order one, we take into account the quadratic source in ``massive" epsilons. Thus, $\dot{\epsilon}_\phi$ can be written as the sum of the general solution to its homogeneous equation, plus a particular solution to Eq. (\ref{dilper}). The former is of order $1/a_0^{D-1}$ and turns out to be dominated at late times by the latter. Actually,  using (\ref{Ptbg14}}), the quadratic source term involves products of Bessel functions with arguments $\lambda_P\, t^{2/D}$ and $\lambda_Q\, t^{2/D}$. Integrating it once, the dominant contribution to $a_0^{D-1}\dot \epsilon_\phi$ is found to arise for ``constructive interferences", \ie when $\lambda_P= \lambda_Q$. This yields the asymptotic behavior,
\be
\dot \epsilon_\phi\sim-{C_\phi\over a_0^{D-2}}\qquad \Longrightarrow\qquad \epsilon_\phi\propto {1\over t^{1-4/D}}\;\mbox{ for } \; D\ge 5\quad \and \quad \epsilon_\phi\propto \ln t \;\mbox{ for }D=4 ,
\ee
where $C_\phi$ is a fully determined coefficient quadratic in $C_{\pm}^M$'s and positive. For $D\ge 5$, the consistency condition $\abs\epsilon_\phi\abs \ll 1$ is automatically fulfilled at late times. On the contrary, the case $D=4$ yields formally to a logarithmically decreasing $\epsilon_\phi$ and one may worry that the our expansions breaks down. Therefore, we have solved numerically the full non-linear differential system (\ref{stab9})--(\ref{stab12}) in this case and found that the perturbative analysis gives the correct late time behavior, which we summarize at the end of this section.

To analyze the evolution of the scale factor and temperature fluctuations, we expand the energy density and pressure around the background (\ref{radera}) and find from Friedmann's equation (\ref{stab9}) and (\ref{stab1310}),
\begin{align}
\label{eq1}
(D-1)(D-2)\, H_0\, \dot\epsilon_a&={1\over 2} \, F_{MN}\abs_{\v\Phi_0} \dot\epsilon^M\dot \epsilon^N+D\rho\abs_{(T_0,\phi_0,\v\Phi_0)}\epsilon_T-{D-3\over 2}\, \F_{MN}\abs_{(T_0,\phi_0,\v\Phi_0)} \epsilon^M\epsilon^N,\\
\label{eq2}
D(\epsilon_a+\epsilon_T)\, \rho\abs_{(T_0,\phi_0,\v\Phi_0)}&={D-2\over 2}\, \F_{MN}\abs_{(T_0,\phi_0,\v\Phi_0)} \epsilon^M\epsilon^N.
\end{align}
It is then straightforward to solve for $\epsilon_a$, whose asymptotic behavior is again dictated by the source terms in ``constructive interferences" arising from the products $\dot\epsilon^M\dot \epsilon^N$ and $\epsilon^M\epsilon^N$ in (\ref{eq1}) and (\ref{eq2}). The late time scaling property of $\epsilon_a$ is found to be
\be
\label{sola}
\epsilon_a\propto {a_0^2\over t}\propto {1\over t^{1-4/D}},
\ee
which can be used in Eq. (\ref{eq2}) to find
\be
\label{solT}
\epsilon_T\propto {a_0^2\over t}\, (1+\mbox{oscillations with constant amplitude}).
\ee
In (\ref{sola}) and (\ref{solT}), the coefficients of proportionality are again fully expressed in terms of the $C_\pm^M$'s.

We signal  that for $D\ge 5$, all terms we have neglected in the perturbed equations of motion are a posteriori found  to be dominated by the sources we took into account. This guarantees the validity of the asymptotic behaviors we have found for the deviations defined in (\ref{stab18}). These results have been confirmed by direct numerical analysis of the unperturbed system of differential equations in some examples.
Since all fluctuations converge to zero, the late time cosmology is radiation dominated. In particular, the dilaton motion  and the damped oscillations of $\t \epsilon^M$ store a negligible amount of energy as compared to the thermal radiation energy. The internal moduli are dynamically stabilized and their effective time-dependent masses  (measured in Einstein frame) are $M_{\t \Phi^M}\propto T_0^{D-2\over 2}e^{2\phi_0\over D-2}$.

For $D=4$, the numerical simulations show that the internal moduli converge to $\v \Phi_0$, while the dilaton decreases logarithmically with time. Individually, the energy stored in the dilaton motion, the total energy (kinetic plus potential) of the damped oscillations of $\v\Phi$, and the thermal radiation energy decay at the same rate. Their late time behavior satisfies
\be
 H^2\propto \dot\phi^2\propto \left(\demi\,  F_{MN}\dot\Phi^M\dot\Phi^N+\rho\right)\propto {1\over a^4},
 \ee
so that the metric evolution is identical to that of a radiation dominated universe, $a\propto \sqrt{t}$.

The above logarithmic behavior of the heterotic dilaton is transferred by heterotic/type I duality to the type I dilaton for $D=4$.  Moreover, in any dimension, stabilization of the internal moduli on the heterotic side implies stabilization of internal moduli on the type I side, except for the special case of $D=6$, where S-duality exchanges the six-dimensional heterotic coupling with the type I internal volume modulus.


\section{Example : Dual heterotic/type I strings on {\boldmath $T^2$}}
\label{example}

Our aim is to illustrate the analysis of the previous section with examples for $D=8$. We want to find local attractor solutions of the form (\ref{radera}) associated to enhanced symmetry points $\v \Phi_0$ of the internal moduli space of the heterotic string on $T^2$. We shall see that the one-loop free energy density has enough structure to stabilize
$\T=B_{89}+i\sqrt{\h g_{88}\h g_{99}-\h g_{89}^2}$, $\U=\big(\h g_{89}+i\sqrt{\h g_{88}\h g_{99}-\h g_{89}^2}\big)/\h g_{88}$ and the Wilson lines $Y_i^I$ ($i,j=8,9$; $I=10,11,\dots,25$). This translates in the type I side into expectation values of the closed and open string internal moduli via the duality map $\T=\T_\rmI$, $\U=\U_\rmI$, $Y_i^I=Y_{\rmI i}^I$, where
\be
\T_{\rm I}= C_{89}+i{\sqrt{\h g_{\rmI 88}\h g_{\rmI 99}-\h g_{\rmI 89}^2}\over \lambda_\rmI}=C_{89}+ie^{-\phi_\rmI}{\left(\h g_{\rmI 88}\h g_{\rmI 99}-\h g_{\rmI 89}^2\right)^{1/4}\over 2\pi}\; , \quad \U_\rmI={\h g_{\rm I 89}+i\sqrt{\h g_{\rm I 88}\h g_{\rm I 99}-\h g_{\rm I 89}^2}\over \h g_{\rm I 88}} .
\ee
The only remaining flat direction of the thermal effective potential corresponds to the heterotic and type I dilatons in eight dimensions, which are related as : $\phi_\rmI= -{1\over 2}\phi-{3\over 4}\ln\big((2\pi)^2 \sqrt{\h g_{88}\h g_{99}-\h g_{89}^2}\big)$.

The heterotic effective action in the Einstein frame is (see for instance appendices D and E in \cite{Kiritsis})
\be
\label{stab1}
    S=\int d^8x \sqrt{-g} \left\{\left[\frac{R}{2}-\frac{(\partial\phi)^2}{3}-\frac{1}{4}\left(\frac{|\partial\cu|^2}{\cu_2^2}+\frac{|\partial\ct+Y^I_{[8}\partial Y^I_{9]}|^2}{\ct_2^2}+\frac{|\cu\partial Y_8^I-\partial Y_9^I|^2}{\ct_2\, \cu_2}\right)\right]-\cf\right\}.
\ee
Indeed,  if we arrange the thirty-four entries of the moduli vector as $\v \Phi\equiv (\ct_1,\ct_2,\cu_1,\cu_2,Y_8^I,Y_9^{I'})$, where indices 1 and 2 refer to real and imaginary parts, the metric components  of the general expression (\ref{stab81}) are
\begin{align}\label{CoMat}
    \left(F_{MN}\right)=\left(
             \begin{array}{cccccc}
               \frac{1}{2\ct_2^2} & 0 & 0 & 0 & -\frac{Y_9^J}{4\ct_2^2} & \frac{Y_8^{J'}}{4\ct_2^2} \\
                & \frac{1}{2\ct_2^2} & 0 & 0 & 0 & 0 \\
                &  & \frac{1}{2\cu_2^2} & 0 & 0 & 0 \\
                &  &  & \frac{1}{2\cu_2^2} & 0 & 0 \\
                & \text{sym.} &  &  & \frac{|\cu|^2}{2\ct_2\cu_2}\delta^{IJ}+\frac{Y_9^IY_9^J}{8\ct_2^2} & -\frac{\cu_1}{2\ct_2\cu_2}\delta^{IJ'}-\frac{Y_9^IY_8^{J'}}{8\ct_2^2} \\
                &  &  &  &  & \frac{1}{2\ct_2\cu_2}\delta^{I'J'}+\frac{Y_8^{I'}Y_8^{J'}}{8\ct_2^2}
             \end{array}
           \right).
\end{align}
The free energy density $\F$  is determined by the mass spectrum (see Eq. (\ref{freegene})), which is specified by the left (right)-moving oscillator number $A$ ($\b A$), the internal momenta and winding numbers $m_i$, $n^i$ ($i=8,9$), and the root vector $Q^I$ of the right-moving internal lattice $\Gamma_{Spin(32)/{\Z_2}}$. As reviewed in the appendix, the mass formula $\hat M^2_s=2(A+\bar{A}) + \demi\left(\vec{p}_L^2+\vec{p}_R^2\right)$ involves the left and right-moving momenta along the compact directions,
\be
\label{stab3}
\begin{array}{ll}
  \dis  p^I_{L,R}=\Big(m_{i}-Q^JY^J_{i}-n^{j} B_{ij}-\demi n^{j}Y^J_{i}Y^J_{j}\Big)e^{*i I} \mp n^{i}e^I_{i}&\!\text{for }\;  i,j,I=8,\dots,9;\;  J=10,\dots,25, \\
    p^I_R=\sqrt{2}\left(Q^I+n^{i}Y^I_{i}\right)&\!\text{for }\; I=10,\dots,25;\; \vec{Q}\in\Gamma_{Spin(32)/\Z},\phantom{\v{\v{\v{\Phi}}}}
\end{array}
\ee
where $\h g_{ij}=e^I_ie^I_j$ and ${e^*}^{iI}e^I_j=\delta^i_j$. More explicitly, one obtains
\be
\label{stab7}
\hat{M}^2_{A,\vec{m},\vec{n},\vec{Q}}(\ct,\cu,Y)=\frac{1}{\ct_2\cu_2}\left|-m_8 \cu+m_9+\tdt n^8+\Big(\tdt \cu-\demi \cw^I \cw^I\Big)n^9+\cw^IQ^I\right|^2+4A,
\ee
where we have defined
\be
\label{stab5}
    \cw^I:=\cu Y^I_8-Y^I_9\; , \qquad \tdt:=\ct+\demi Y^I_8\cw^I
\ee
and used the level matching condition, $A-\bar{A}=m_in^i+\demi Q^IQ^I$. At generic points in moduli space, the gauge symmetry is $U(1)^2_L\times U(1)^2_R\times U(1)^{16}_R$, where $U(1)^2_L\times U(1)^2_R$ arises from $T^2$ compactification, and  $U(1)^{16}_R$ is the Cartan subgroup of $SO(32)_R$. We now examine special points in moduli space where $n_0$ pairs of bosonic and fermionic superpartners generically massive are accidentally massless. Since at zero temperature the model is maximally supersymmetric, such points are associated to enhanced gauge symmetries. In fact, the additional massless modes arise at oscillator levels $A=0$, $\b A=-1$, so that $n_0$ is proportional to $s_0r_{-1}=2^3$ (see the appendix) and the enhancements of the gauge theory arise from the right-moving sector only. In the following two examples, we will simplify the notations by omitting the subscript ``$R$'' in the right-moving gauge group factors.

\vspace{-.3cm}
\subsection*{\it Local attractor 1 : $U(1)_L^2\times SU(3) \times SO(32)$ }
\vspace{-.3cm}
We start with the most obvious attractor where all Wilson lines vanish, $Y^I_i=0$, leaving the $SO(32)$ group unbroken.  The torus moduli take the values $\ct=\cu=\demi+i\frac{\sqrt{3}}{2}$, implying an additional $SU(3)$ gauge factor. The $n_0$ states responsible for the enhancement of $U(1)^2\times U(1)^{16} \rightarrow SU(3) \times SO(32)$ are divided into two groups :

$\bullet$ $6\times 2^3$ boson/fermion pairs imply $U(1)^2\rightarrow SU(3)$.  Their quantum numbers are $(\vec{m},\vec{n})=\pm(1,1;0,1)$, $\pm(0,1;-1,1)$ or $\pm(1,0;1,0)$, and $\vec{Q}=0$. In this case, $p_L^{8,9}=0=p_R^{I\geq 10}$ and $(p^8_R,p^9_R)$ realize the root vectors of $SU(3)$, which represent a hexagon. The corresponding 6 mass formulas are,
\be
\label{massSU3}
            \hat{M}_{0,\vec{m}, \vec{n}, \v 0}^2=\left\{
            \begin{array}{ll}
                \frac{1}{\ct_2\cu_2}|1-\cu+\tdt\cu -\demi \cw^I \cw^I |^2&\text{for }(\vec{m},\vec{n})=\pm(1,1,0,1),\\
                \frac{1}{\ct_2\cu_2}|1-\tdt+\tdt\cu -\demi \cw^I \cw^I |^2&\text{for }(\vec{m},\vec{n})=\pm(0,1,-1,1),\\
                \frac{1}{\ct_2\cu_2}|\tdt-\cu|^2& \text{for }(\vec{m},\vec{n})=\pm(1,0,1,0).
            \end{array}
                \right.
\ee

$\bullet$ $480\times 2^3$ boson/fermion pairs to recover $U(1)^{16}\to SO(32)$. They have $(\vec{m},\vec{n})=(\v 0,\v 0)$, $\vec{Q}=\pm(1,\pm 1, 0, ..., 0)$, $\pm(1,0, \pm 1, ..., 0)$ or any other permutation. In this case, $p_L^{8,9}=0=p_R^{8,9}$, while $(p_R^{I\geq10})$ realize the root vectors of $SO(32)$. The corresponding 480 mass formulas are
\be\label{mSO32}
    \hat{M}_{0, \v0, \v0, \v Q}^2=\frac{1}{\ct_2\cu_2}\left|\pm(\cw^I\pm\cw^J)\right|^2\; , \qquad I,J=10,\dots ,25\; , \ I\neq J.
\ee
To compute the squared mass matrix defined in Eq. (\ref{Ptbg10}), we first evaluate the second derivatives (\ref{2dder}) of the free energy  at $\v\Phi_0$.  The non vanishing components are proportional to
\be
\label{stab22}
\!\!\!\begin{array}{ll}
\dis \left.\sum_{u=1}^{n_0}\frac{\partial^2\hat{M}^2_u}{\partial \ct_\alpha \partial \ct_\alpha}\rst_{\v\Phi_0}\!\!=\left.\sum_{u=1}^{n_0}\frac{\partial^2\hat{M}_u^2}{\partial \cu_\alpha \partial \cu_\alpha}\rst_{\v\Phi_0}\!\!=16\times 2^3\!\!\!& ,\; \alpha=1,2 \; \text{(no sum over }\alpha)\\
\dis \left.\sum_{u=1}^{n_0}\frac{\partial^2\hat M^2_u}{\partial Y_i^I \partial Y_i^I}\rst_{\v \Phi_0}\!\!=-2  \left.\sum_{u=1}^{n_0}\frac{\partial^2\hat M^2_u}{\partial Y_8^I \partial Y_9^I}\rst_{\v\Phi_0}\!\!=160\times 2^3\!\!\!&
 ,\; i=8,9; \;  I=10,\dots ,25 \; \text{(no sum over }i, I).
\end{array}
\ee
The nonzero entries of the metric (\ref{CoMat}) at $\v \Phi_0$ are also found to be
\be
\label{stab24}
\begin{array}{ll}
\dis F_{\ct_\alpha\ct_\alpha}=F_{\cu_\alpha\cu_\alpha}=\frac{2}{3}& ,\; \alpha=1,2 \; \text{(no sum over }\alpha)\\
\dis F_{Y^I_iY^I_i}=\frac{2}{3}\; , \;\; \;\;F_{Y^I_8Y^I_9}=-\frac{1}{3} &,\;
 i=8,9; \;  I=10,\dots ,25 \; \text{(no sum over }i, I).
\end{array}
\ee
The resulting matrix of squared masses is diagonal, with strictly positive eigenvalues. Therefore, all flat directions of the internal moduli space are lifted. Once the dynamics is attracted to the trajectory (\ref{radera}), the ``time-dependent moduli squared masses" are
\be
\label{stab26}
    M_{\Phi_1}^2 =  \frac{c_6}{4 \pi} \, 2^3 \times 24 \, e^{\frac{2\phi_0}{3}} \, T_0^{6} \qquad \mbox{ or }\qquad M_{\Phi_2}^2 = \frac{c_6}{4 \pi} \, 2^3 \times 240 \, e^{\frac{2\phi_0}{3}} \, T_0^{6} \, .
\ee
The first one corresponds to $\ct_1$, $\ct_2$, $\cu_1$, $\cu_2$, while the second is associated to the Wilson lines $Y_8^I$ and $Y_9^I$. The additional factor of ten for the latter can be understood from the fact that they are coupled to ten times as many additional states as compared to the torus moduli.

\vspace{-.3cm}
\subsection*{\it Local attractor 2 : $U(1)_L^2\times SU(2) \times SO(34)$}
\vspace{-.3cm}
The point $\v \Phi_0$ we now consider corresponds to the values $\ct=\cu=i/\sqrt{2}$, $Y^{I\ge 10}_8=0$ and $Y^{10}_9=-Y^{11}_9 = -Y^{12}_9 = \dots =- Y^{25}_9 =-1/2$. This moduli configuration is much less trivial than the previous one, since it is going to give rise to the gauge group $SU(2)_{8} \times SO(34)_{9,\dots,25}$, where the subscripts denote which directions $i = 8,9$ and $I=10,...,25$ are associated with the gauge factors.  There are $n_0=546\times 2^3$ extra massless boson/fermion pairs of states, which can be divided into  $2\times 2^3$ for the  $SU(2)_{8}$ and $544\times 2^3$ for the $SO(34)_{9,\dots,25}$ enhancements.  Note that the $SO(34)_{9,\dots,25}$ factor arises from an enhancement of the $U(1)_{9}$ symmetry of the $T^2$ torus, with the $SO(32)$ symmetry of the internal lattice.  The detailed quantum numbers of the extra states are as follows :

$\bullet$ $2\times 2^3$ boson/fermion pairs give $U(1)_{8}\rightarrow SU(2)_{8}$. They have $(\vec{m},\vec{n})=\pm(1,0;1,0)$ and $\vec{Q}=0$. In this case, $p_L^{I\ge 8}=0=p_R^{J\geq 9}$, while $p^8_R=\pm\sqrt{2}$ realize the root vectors of $SU(2)_8$.

For $SO(34)_{9,\dots,25}$, the $544\times 2^3$ pairs of bosons and fermions giving $U(1)^{17}_{9,\dots,25} \rightarrow SO(34)_{9,\dots,25}$ are subdivided into :

$\bullet$ $420\times 2^3$ pairs transform in the adjoint representation of $SO(30)$ and are giving rise to $U(1)^{15}_{11,\dots,25}\rightarrow SO(30)_{11,\dots,25}$. $210\times 2^3$ have $(\vec{m},\vec{n},\vec{Q})=\pm(0,1;0,0;0,1,1,0,...,0)$ or any permutation of the last 15 entries. The other $210\times 2^3$ have $(\vec{m},\vec{n},\vec{Q})=(0,0;0,0;0,1,-1,0,...,0)$ or any  permutation of the last 15 entries.

$\bullet$ $60\times 2^3$ pairs transform as $(2,30)$ under $SO(2)_{10} \times SO(30)_{11,\dots,25}$, giving the enhanced group $SO(32)_{10,\dots,25}$. $30\times 2^3$ of them have $(\vec{m},\vec{n},\vec{Q})=\pm(0,1;0,0;-1,1,0,...,0)$ or any  permutation of the last 15 entries. The other $30\times 2^3$ have $(\vec{m},\vec{n},\vec{Q})=\pm(0,0;0,0;  1 ,1,0,...,0)$ or any permutation of the last 15 entries.

$\bullet$ $64\times 2^3$ pairs transform as $(2,32)$ under $SO(2)_{9}\times SO(32)_{10,\dots25}$, giving the enhanced gauge group $SO(34)_{9,\dots,25}$. $32\times 2^3$ of them have  $(\vec{m},\vec{n},\vec{Q})=\pm(0,1;0,-1;\demi,...,\demi)$ and $\pm(0,1;0,-1;-\demi,-\demi,\demi,...,\demi)$ or any permutation of the last 15 entries. The other $32\times 2^3$ have $(\vec{m},\vec{n},\vec{Q})=\pm(0,2;0,-1;-\frac{3}{2},\demi,...,\demi)$ and $\pm(0,2;0,-1;-\demi,\frac{3}{2},\demi,...,\demi)$ or any permutation of the last 15 entries.

Proceeding as before, the squared mass matrix in (\ref{Ptbg10}) can be evaluated.  Its diagonalization reveals two groups of eigenvalues,
\begin{align}
M_{\tilde \Phi_1}^2 = \frac{c_6}{4 \pi} \, 2^3 \times 16 \, e^{\frac{2\phi_0}{3}} \,  T_0^6 \; , \qquad \qquad M_{\tilde \Phi_2}^2 = \frac{c_6}{4 \pi} \,2^3 \times 256 \, e^{\frac{2\phi_0}{3}} \, T_0^6 \, .
\end{align}
The first one is associated to $\ct_1-\cu_1-{1\over 4}(Y^{10}_8-Y^{11}_8-\cdots-Y^{25}_8)$ and $\ct_2-\cu_2$, while the second corresponds to $\ct_1+\cu_1$, $\ct_2+\cu_2$ and all 32 Wilson lines.  Thus, we find a second point in moduli space where all internal moduli are stabilized by the thermal effective potential\footnote{We have also investigated a third local attractor at the point $\ct=\cu=i/2$, $Y_8^{I\ge 10}=0=Y^{10,11}_9$, $Y^{12,\dots,25}_9=1/2$, which corresponds to the gauge enhancement $SU(2)\times SU(2)\times SO(32)$ and a stabilization of all internal moduli. Due to its similarity, we do not present its details here.}.


\section{Conclusions and perspectives}
\label{conclu}

In this paper, we considered toroidally compactified heterotic and type I superstrings at finite temperature.  Applying the rules of heterotic/type I duality, we inferred novel contributions to the free energy of a gas of type I superstrings.  These contributions are due to BPS D-strings wrapped on internal circles which become massless at special points in moduli space, enhance the gauge group, and lift flat directions.  These conclusions are based on the S-dual heterotic picture at weak coupling.  At finite temperature, the latter is a no-scale model \ie a flat background where all supersymmetries are spontaneously broken at tree level.

We computed the one-loop free energy density on the heterotic side for $D \geq 4$ and found points in moduli space where all internal moduli are dynamically stabilized due to the cosmological evolution.  Additionally, in $D \geq 5$, the evolution of the dilaton asymptotes to a constant value, while in $D=4$, the dilaton turns out to have a logarithmically decreasing behavior.

Using the S-duality, this implies that for $D \geq 7$, all type I internal moduli can be stabilized at strong coupling.  In $D=6$, the S-duality maps the heterotic coupling into the type I volume modulus.  As a result the only remaining flat direction in type I is the internal volume modulus, which asymptotes to a constant finite value, while the type I dilaton is stabilized at weak coupling.  For the cases $D \leq 5$, all type I internal moduli can be stabilized at weak coupling.  Furthermore in $D=4$, the type I dilaton inherits the logarithmic behavior from the heterotic dilaton, while it asymptotes to a constant in higher dimensions.   In all cases, the late time geometric evolution is identical to a radiation dominated evolution.  Furthermore, all solutions are stable under small perturbations and are thus local attractors of the dynamics.

It is worth stressing that the effects of the massless BPS non-perturbative D-strings persist at weak coupling, as their masses are protected by supersymmetry.  As a result, the stabilization in type I for $D\geq 7$ persists at weak coupling.  Furthermore, taking these modes into account is not optional in phenomenologically motivated uses of the type I superstring.  Actually, this is not the first time massless solitons play an essential role in weakly coupled theories. For instance, in type IIB compactifications on Calabi-Yau threefolds, the conifold singularities in the vector multiplets moduli spaces are explained by massless hypermultiplets realized by D3-branes wrapped on vanishing 3-cycles \cite{stro}.

Realistic models should include also a spontaneous breaking of $\N_4=1$ supersymmetry at a scale $M$, before finite temperature $T$ is switched on.  In this case, the universe is attracted to a ``radiation-like dominated era" \cite{cosmoA,cosmoB,cosmoC,cosmoreview,cosmoD,cosmoDb}. This evolution is characterized by coherent motions of $e^{4\phi(t)}$ (where $\phi$ is the dilaton in four dimensions) and the modulus $M(t)$, both proportional to $T(t)$ such that Friedmann's equation is effectively that of a radiation dominated era, $H^2\propto T^4$.  The energy stored in the oscillations of the moduli around their minima is found to be dominated by the thermal energy and so the stabilization of the scalars is guaranteed.  Moreover, infrared effects are expected to put a halt to the run away behavior of the string coupling and supersymmetry breaking scale. In particular, when $T(t)$ reaches the electroweak scale $M_{\rm EW}$, radiative corrections are not screened anymore by temperature effects and the electroweak breaking is expected to take place \cite{AlvarezGaume:1983gj}. This should be accompanied by the stabilization of $M(t)$ around $M_{\rm EW}$ \cite{Kounnas:1994fr}.  Clearly, it is of utmost importance to implement these effects in our cosmological set up since this would provide a precise context for addressing questions of dark matter, astroparticle physics and phenomenology.  Additionally for $D=4$, as well as $D=5$, there is the possibility of large contributions coming from light NS5-brane states in the heterotic theory or D5-brane states in the type I theory which have not been taken into account yet.  It is possible that these states can play a role in stabilizing the dilaton.  To make progress in this direction, one may try to exploit heterotic/type II duality in $D=4$ which is a strong-weak duality.


\section*{Acknowledgements}

We are grateful to C. Bachas, E. Dudas, I. Florakis, C. Kounnas, A. Sagnotti and N. Toumbas for useful discussions. H.P. would like to thank C.E.R.N. where part of this work was completed.

\noindent
J.E. acknowledges financial support from the Groupement d'Int\'er\^et Scientifique P2I, as well as support by the FWO - Vlaanderen, Project No. G.0235.05, and by the ``Federal Office for Scientific, Technical and Cultural Affairs through the Interuniversity Attraction Poles Programme Belgian Science Policy" P6/11-P.
The work of L.L. and H.P. is partially supported by the contracts PITN GA-2009-237920, ERC-AG-226371, ANR 05-BLAN-NT09-573739, CEFIPRA/IFCPAR 4104-2 and PICS France/Cyprus, France/Greece, France/USA.


\section*{Appendix : Thermal partition functions}
\renewcommand{\theequation}{A.\arabic{equation}}
\renewcommand{\thesection}{A}
\setcounter{equation}{0}


\subsection*{\it Type I superstring}
\vspace{-.3cm}
To study the canonical ensemble of a perfect gas of maximally supersymmetric open and closed superstrings, we compactify the type I theory on the Euclidean background $S^1(R_{{\rm I}0})\times T^{D-1}\times \prod_{i=D}^9 S^1(R_{{\rm I}i})$. Bosons (fermions) are imposed periodic (antiperiodic) boundary conditions along $S^1(R_{{\rm I}0})$, where $\h\beta_{\rm I}=2\pi R_{{\rm I}0}$ is the inverse temperature. The spatial torus $T^{D-1}$ is considered in the large volume $\h V_{\rm I}$ limit. Our aim is to compute the one-loop thermal partition function. The treatment of a generic Scherk-Schwarz compactification can be found in \cite{ADS} and the case of present interest is reviewed in \cite{Angelantonj:2002ct}.

In the closed string sector, the torus contribution is half that of type IIB,
\be
\label{ap1}
\begin{array}{rl}
    \T=&\!\!\!\dis {\h \beta_{\rm I} \hat V_{\rm I}\over (2\pi)^D}\,  {1\over 2} \int_{\F}{d^2\tau\over 2  \tau_2^{{D\over 2}+1}}\,  {1\over \eta^8\b\eta^8}\, \sum_{\v m,\v n}q^{{1\over 4}\v p_L^2}\b q^{{1\over 4}\v p_R^2} \sum_{n^0,\t m_0}e^{-\frac{\pi R_{{\rm I}0}}{\tau_2}\lst  n^0\tau+\tilde{m}_0\rst^2}\\
    &\!\!\!\dis \times{1\over 2}\sum_{a,b}(-)^{a+b+ab}\frac{\theta[\substack{a\\b}]^4}{\eta^4}\, {1\over 2}\sum_{\bar{a},\bar{b}}(-)^{\bar{a}+\bar{b}+\bar{a}\bar{b}}\frac{\bar{\theta}[\substack{\bar{a}\\\bar{b}}]^4}{\bar{\eta}^4}\, (-)^{\tilde{m}_0(a+\bar{a})+n^0(b+\bar{b})} \\
    =&\!\!\!\dis {\h \beta_{\rm I} \hat V_{\rm I}\over (2\pi)^D}\,  {1\over 2} \int_{\F}{d^2\tau\over 2  \tau_2^{{D\over 2}+1}}\,  {1\over \eta^8\b\eta^8}\, \sum_{\v m,\v n}q^{{1\over 4}\v p_L^2}\b q^{{1\over 4}\v p_R^2} \\
    &\!\!\!\dis \Bigg\{\sum_{n^0 \, {\rm even}, \, \t m_0}e^{-{\pi R^2_{{\rm I}0}\over \tau_2}\abs n^0\tau+\t m_0\abs^2}\Big[(V_8\b V_8+S_8\b S_8)-(-1)^{\t m_0}(V_8\b S_8+S_8\b V_8)\Big]\\
    &\!\!\!\dis \;\;+\!\!\sum_{n^0\,  {\rm odd}, \, \t m_0}e^{-{\pi R^2_{{\rm I}0}\over \tau_2}\abs n^0\tau+\t m_0\abs^2}\Big[(O_8\b O_8+C_8\b C_8)-(-1)^{\t m_0}(O_8\b C_8+C_8\b O_8)\Big]\Bigg\},
\end{array}
\ee
where $q=e^{2i\pi \tau}$ and $p_{L,Ri}=m_i/R_{{\rm I}i}\mp n^i R_{{\rm I}i}$. The above second expression involves $SO(8)$ affine characters, where those associated to the vectorial and spinorial representations satisfy
\be
\label{V=S}
    {V_8\over \eta^8}={S_8\over \eta^8}=\sum_{A\ge 0}s_A\, q^A.
\ee
The Klein bottle amplitude $\K$ is obtained by keeping all characters of $\T$ which are invariant under left $\leftrightarrow$  right symmetry. Symmetrizing and antisymmetrizing the NS-NS and RR sectors respectively, $\K$ involves the combination $V_8-S_8$ and is thus vanishing.
In the open string sector, the thermal annulus and M\"obius strip amplitudes are
\begin{eqnarray}
\label{ap3}
\A &\!\!\!=\!\!\!&{\h \beta_{\rm I} \hat V_{\rm I}\over (2\pi)^D}\,  {N^2\over 2} \int_0^{+\infty} {d\tau_2\over 2\tau_2^{{D\over 2}+1}}\, {1\over \eta^8}\, \sum_{\v m} q^{\v p^2}\, \sum_{\tilde m_0} e^{-{\pi R_{{\rm I}0}^2\over \tau_2}\t m_0^2}\Big[ V_8-(-1)^{\tilde m_0}S_8\Big],\\
\label{ap4}
\M &\!\!\!=\!\!\!& -{\h \beta_{\rm I} \hat V_{\rm I}\over (2\pi)^D}\,  {N\over 2} \int_0^{+\infty} {d\tau_2\over 2\tau_2^{{D\over 2}+1}}\, {1\over\h \eta^8}\, \sum_{\v m} q^{\v p^2}\, \sum_{\tilde m_0} e^{-{\pi R_{{\rm I}0}^2\over \tau_2}\t m_0^2}\Big[ \h V_8-(-1)^{\tilde m_0}\h S_8\Big],
\end{eqnarray}
where $N=32$, $q=e^{-\pi \tau_2}$, $p_i=m_i/R_{{\rm I}i}$ and the ``hatted" characters in Eq. (\ref{ap4}) have the power expansion
\be
\label{ap5}
{\h V_8\over\h\eta^8}={\h S_8\over\h\eta^8}=\sum_{A\ge 0} (-)^As_A\, q^A.
\ee

We proceed by evaluating more explicitly  the amplitude $\T$ by ``unfolding" the fundamental domain of integration \cite{McClain:1986id}.   In fact, for any set of modular covariant functions $f_{(n,\t m)}(\tau,\b \tau)$ such that $f_{(n,\t m)}(M(\tau),M(\b \tau))=f_{(n,\t m)M}(\tau,\b \tau)$ for all $M\in SL(2,\Z)$, one has\footnote{Eq. (\ref{unfo}) is true as long as it is allowed to exchange discrete sum and integration, a fact which is guaranteed if the integrand is absolutely convergent. This condition is satisfied for $\T$ when $R_{{\rm I}0}>R_{{\rm I}{\rm H}}$.}
\be
\label{unfo}
\int_\F{d^2\tau\over \tau_2^2}\, \sum_{n,\t m}f_{(n,\t m)}(\tau,\b \tau)=\int_\F{d^2\tau\over \tau_2^2}\, f_{(0,0)}(\tau,\b \tau)+\int_{\S_+}{d^2\tau\over \tau_2^2}\, \sum_{\t m\neq 0}f_{(0,\t m)}(\tau,\b \tau),
\ee
where $\S_+$ is the upper half strip : $-1/2< \tau_1< 1/2$, $\tau_2>0$. Applied to Eq. (\ref{ap1}), supersymmetry implies that the contribution for $n_0=\t m_0=0$ vanishes and we are left with integrals  over $\S_+$ for $n_0=0$, $\t m_0\neq 0$. Defining $\t m_0=2\t k_0+1$ and using (\ref{V=S}), one obtains
\begin{eqnarray}
    \T&\!\!\!=\!\!\!&{\h \beta_{\rm I} \hat V_{\rm I}\over (2\pi)^D} \int_{\S_+}{d^2\tau\over \tau_2^{{D\over 2}+1}}\sum_{\scriptsize \substack{\t k_0,\, \v m, \, \v n\\A\ge 0,\, \b A \geq 0}}s_As_{\b A} \,e^{2i\pi \tau_1(A-\b A-\v m\cdot\v n)}\, e^{-{\pi R_{{\rm I}0}^2\over\tau_2}(2\t k_0+1)^2-\pi \tau_2\big[2(A+\b A)+\sum_i\big(\frac{m_i^2}{R_{{\rm I}i}^2}+{n^i}^2R_{{\rm I}i}^2 \big) \big]}\nonumber\\
\label{ap7}
&\!\!\!=\!\!\!&{\h \beta_{\rm I} \hat V_{\rm I}\over (2\pi)^D} \int_0^{+\infty}{d\tau_2\over \tau_2^{{D\over 2}+1}}\sum_{\scriptsize \substack{\t k_0, \,  \v m, \, \v n\\ A\ge 0,\,  \b A\geq 0\\ A-\b A = \v m\cdot \v n}}s_As_{\b A}e^{-{\pi R_{{\rm I}0}^2\over\tau_2}(2\t k_0+1)^2-\pi \tau_2\big[ 4A +\sum_i\big(\frac{m_i}{R_{{\rm I}i}}-n^iR_{{\rm I}i} \big)^2 \big]},
\end{eqnarray}
where level matching is implemented by integrating over $\tau_1$.
Using the formula $\int_0^{\infty}dx\frac{e^{-a/x-bx}}{x^{\nu}}= 2a^{\frac{1-\nu}{2}} b^{\frac{\nu-1}{2}} K_{\nu-1} (2\sqrt{ab})$, where $K_{\nu}(x)$ is the Bessel function of second kind, the integral over $\tau_2$ yields Eqs. (\ref{ZT}) and (\ref{cg}). Similarly, applying the expansions (\ref{V=S}) and (\ref{ap5}) in Eqs. (\ref{ap3}) and (\ref{ap4}), we have
\begin{align}\label{ap8}
    \A =&{\h \beta_{\rm I} \hat V_{\rm I}\over (2\pi)^D}\,  {N^2\over 2} \int_0^{+\infty} {d\tau_2\over \tau_2^{{D\over 2}+1}} \sum_{\tilde k_0,\, \v m , \, A\geq0}s_A \, e^{-{\pi R_{{\rm I}0}^2\over \tau_2}(2\t k_0+1)^2-\pi\tau_2\big(\sum_i{m_i^2\over R_{{\rm I}i}^2}+A\big)},\\
\label{ap9}
    \M =&-{\h \beta_{\rm I} \hat V_{\rm I}\over (2\pi)^D}\,  {N\over 2} \int_0^{+\infty} {d\tau_2\over \tau_2^{{D\over 2}+1}} \sum_{\tilde k_0,\, \v m , \, A\geq0}(-)^A s_A \, e^{-{\pi R_{{\rm I}0}^2\over \tau_2}(2\t k_0+1)^2-\pi\tau_2\big(\sum_i {m_i^2\over R_{{\rm I}i}^2}+A\big)},
\end{align}
which gives Eq. (\ref{ZA}) after integration over $\tau_2$.


\vspace{-.3cm}
\subsection*{\it Dual heterotic string}
\vspace{-.3cm}
We proceed by deriving the partition function of the dual heterotic theory, which is compactified on $S^1(R_{{\rm h}0})\times T^{D-1}\times \prod_{i=D}^9 S^1(R_{{\rm h}i})$. Bosons and fermions are again given periodic and antiperiodic boundary conditions along the Euclidean time circle, whose circumference defines the inverse temperature $\h\beta_{\rm h}=2\pi R_{{\rm h}0}$. This yields
\begin{align}
\label{ZhApp1}
\dis Z_{\rm h}=&\;{\h \beta_{\rm h} \hat V_{\rm h}\over (2\pi)^D} \int_{\F}{d^2\tau\over 2\tau_2^{{D\over 2}+1}}\, {\Gamma_{(0,16)}\over \eta^8\b\eta^{24}}\, \sum_{\v m,\v n}q^{{1\over 4}\v p_L^2}\b q^{{1\over 4}\v p_R^2}\nnR
    &\;\;\;\;\;\;\;\;\;\;\;\;\;\;\;\;\;\;\;\;\;\;\;\;\;\;\;\;\;\;\;\;\;\,\times \sum_{n^0,\t m_0}e^{-{\pi R_{{\rm h}0}^2\over \tau_2}\abs n^0\tau+\t m_0\abs^2}\, {1\over 2}\sum_{a,b}(-)^{a+b+ab}{\theta[\substack{a\\b}]\over \eta^4}\, (-)^{\t m_0 a+n^0 b+\t m_0 n^0}\nnR
    =&\;{\h \beta_{\rm h} \hat V_{\rm h}\over (2\pi)^D} \int_{\F}{d^2\tau\over 2 \tau_2^{{D\over 2}+1}}\, {\Gamma_{(0,16)}\over \eta^8\b\eta^{24}}\, \sum_{\v m,\v n}q^{{1\over 4}\v p_L^2}\b q^{{1\over 4}\v p_R^2}\, \Bigg\{\sum_{n^0\, {\rm even},\, \t m_0}e^{-{\pi R_{{\rm h}0}\over \tau_2}\abs n^0\tau+\t m_0\abs^2}\Big[V_8-(-1)^{\t m_0}S_8\Big]\nnR
    & \dis\;\;\;\;\;\;\;\;\;\;\;\;\;\;\;\;\;\;\;\;\;\;\;\;\;\;\;\;\;\;\;\;\;\;+\sum_{n^0\, {\rm odd},\, \t m_0}e^{-{\pi R_{{\rm h}0}\over \tau_2}\abs n^0\tau+\t m_0\abs^2}\Big[(-1)^{\t m_0}O_8-C_8\Big]\Bigg\},
\end{align}
where $q=e^{2i\pi\tau}$, while $p_{L,Ri}=m_i/R_{{\rm h}i}\mp n^i R_{{\rm h}i}$ and the volume $\h V_{\rm h}$ are now measured in the heterotic theory. Alternatively, the lattice of internal zero modes can be considered in its Lagrangian formulation, as needed in  section \ref{E1} for the direction 9,
\be
\label{poisson}
\sum_{m_9,n_9}q^{{1\over 4}p_{L9}^2}\b q^{{1\over 4}p_{R9}^2}={R_{{\rm h}9}\over \sqrt{\tau_2}}\sum_{n^9,\t m_9}e^{-{\pi R_{{\rm h}9}^2\over \tau_2}\abs n^9\tau+\t m_9\abs}.
\ee
To unfold the fundamental domain of integration in (\ref{ZhApp1}), one can use the identity (\ref{unfo}) as in the torus amplitude in type I. Expanding the $SO(32)$ right-moving lattice as
\be
\label{ap10}
    {\Gamma_{(0,16)}\over \b\eta^{24}}=\sum_{\b A\ge -1} b_{\b A}\, \b q^{\b A},
\ee
and using Eq. (\ref{V=S}), one obtains
\begin{align}
    \dis Z_{\rm h}=&\; {\h \beta_{\rm h} \hat V_{\rm h}\over (2\pi)^D} \int_{\S_+}{d^2\tau\over \tau_2^{{D\over 2}+1}}\sum_{\scriptsize \substack{\t k_0,\,  \v m, \, \v n\\A\geq0,\, \b A \geq -1}}s_Ab_{\b A} \, e^{2i\pi \tau_1(A-\b A-\v m\cdot\v n)}e^{-{\pi R_{{\rm h}0}^2\over\tau_2}(2\t k_0+1)^2-\pi \tau_2\big[2(A+\b A)+\sum_i\big(\frac{m_i^2}{R_{{\rm h}i}^2}+{n^i}^2R_{{\rm h}i}^2 \big) \big]}\nnR
    =&\;{\h \beta_{\rm h} \hat V_{\rm h}\over (2\pi)^D} \int_0^{\infty}{d\tau_2\over \tau_2^{{D\over 2}+1}}\sum_{\scriptsize \substack{\t k_0, \,  \v m, \, \v n\\A\geq0,\, \b A\geq -1\\ A-\b A=\v m\cdot \v n}}s_Ab_{\b A}\, e^{-{\pi R_{{\rm h}0}^2\over\tau_2}(2\t k_0+1)^2-\pi \tau_2\big[ 4A +\sum_i\big(\frac{m_i}{R_{{\rm h}i}}-n^iR_{{\rm h}i} \big)^2 \big]}, \label{ZhApp}
\end{align}
which can be integrated to give Eqs. (\ref{Zh2}) and (\ref{cg}).


\vspace{-.3cm}
\subsection*{\it Heterotic string at generic point in moduli space}
\vspace{-.3cm}
In sections \ref{stab} and \ref{example} for $D=8$, we study in the context of the maximally supersymmetric heterotic string the stabilization of all internal moduli by the free energy density at weak coupling. In Einstein frame, the latter is $\F=-e^{{2D\over D-2}\phi} Z_{\rm h}/(\h\beta_{\rm h} \h V_{\rm h})$, where $\phi$ is the dilaton in dimension $D$ and $Z_{\rm h}$ is the vacuum energy in the Euclidean background $S^1(R_{{\rm h}0})\times T^{D-1}\times T^{10-D}$.  The internal moduli are the metric $\h g_{ij}$, the antisymmetric tensor $B_{ij}$ and the Wilson lines $Y^I_i$ ($i,j=D,\dots,9$; $I=10,11,\dots,25$). Proceeding as before, the partition function $Z_{\rm h}$ takes the following forms,
\begin{align}
    Z_{\rm h}=&\; {\h \beta_{\rm h} \hat V_{\rm h}\over (2\pi)^D} \int_{\F}{d^2\tau\over 2 \tau_2^{{D\over 2}+1}}\!\sum_{\v m,\v n,\, \v Q}\!{q^{{1\over 4}\v p_L^2}\b q^{{1\over 4}\v p_R^2}\over \eta^8\b\eta^{24}}\sum_{n^0,\t m_0}e^{-{\pi R_{{\rm h}0}^2\over \tau_2}\abs n^0\tau +\t m_0\abs^2}\, {1\over 2}\sum_{a,b}(-)^{a+b+ab}{\theta[\substack{a\\b}]\over \eta^4}\, (-)^{\t m_0 a+n^0 b+\t m_0 n^0}\nnR
    =&\; {\h \beta_{\rm h} \hat V_{\rm h}\over (2\pi)^D}\int_{\cs_+}{d^2\tau\over 2 \tau_2^{{D\over 2}+1} }\sum_{\v m,\v n,\, \v Q} q^{\frac{1}{4}\vec{p}_L^2}\bar{q}^{\frac{1}{4}\vec{p}_R^2} \sum_{\tilde{k}}e^{-\frac{\pi R_{{\rm h}0}^2}{\tau_2}(2\tilde{k}_0+1)^2}\frac{V_8+S_8}{\eta^8\bar{\eta}^{24}} \label{het1}\\
    =&\; {\h \beta_{\rm h} \hat V_{\rm h}\over (2\pi)^D}\int_{\cs_+}\frac{d^2\tau}{\tau^{{D\over2}+1}_2} \sum_{\scriptsize\substack{\tilde{k}_0,\, \v m,\, \v n, \, \v Q\\A\geq0,\, \bar{A}\geq-1}}s_Ar_{\bar{A}}\, e^{2i\pi  \tau_1\left(A-\bar{A}+{1\over 4}(\v p_L^2-\v p_R^2)\right)}  \,  e^{-\frac{\pi R_{{\rm h}0}^2}{\tau_2 }(2\tilde{k}_0+1)^2-\pi\tau_2\left[2(A+\bar{A}) + {1\over 2}(\v p_L^2+ \v p_R^2)\right]},\nonumber
\end{align}
where we introduce the coefficients $r_{\b A}$ of the expansion $\b\eta^{-24}=\sum_{\b A\geq-1}r_{\b A}\b q^{\b A}$. The moduli-dependent internal momenta are specified by $\v m$, $\v n$ and the root vector $Q^I$ of the right-moving lattice $\Gamma_{Spin(32)/\Z_2}$ \cite{HComp},
\be
\label{het2}
\begin{array}{ll}
  \dis  p^I_{L,R}=\Big(m_{i}-Q^JY^J_{i}-n^{j} B_{ij}-\demi n^{j}Y^J_{i}Y^J_{j}\Big)e^{*i I} \mp n^{i}e^I_{i}&\!\!\text{for }  i,j,I=D,\dots,9;\;  J=10,\dots,25, \\
    p^I_R=\sqrt{2}\left(Q^I+n^{i}Y^I_{i}\right)&\!\!\text{for } I=10,\dots,25;\; \vec{Q}\in\Gamma_{Spin(32)/\Z},\phantom{\v{\v{\v{\Phi}}}}
\end{array}
\ee
where $\{e_i\}$ is a vector basis of $T^{10-D}$ \ie $\h g_{ij}=e^I_ie^I_j$ and ${e^*}^{iI}e^I_j=\delta^i_j$. Since these momenta satisfy $\demi(\v p_L^2-\v p_R^2)=-2\v m\cdot\v n-\v Q\cdot\v Q$, the level matching condition implemented by integrating over $\tau_1$ in Eq. (\ref{het1}) is $A-\b A=\v m\cdot\v n+\demi\v Q\cdot\v Q$, which yields
\be
\label{het5}
    Z_{\rm h}   ={\h \beta_{\rm h} \hat V_{\rm h}\over (2\pi)^D}\int_0^{\infty}\frac{d\tau_2}{\tau^{{D\over2}+1}_2}\!\! \sum_{\scriptsize\substack{\t k_0,\, \v m, \, \v n,\, \v Q\\ A\geq0,\, \bar{A}\ge -1\\A-\b A=\v m\cdot\v n+\demi \v Q\cdot\v Q}}\!\! s_Ar_{\bar{A}}\,
   e^{-{\pi R_0^2\over\tau_2 }(2\t k_0+1)^2-\pi\tau_2 \h M_{A,\v m,\v n,\v Q}^2(\h g,B,Y)},
\ee
where $\h M_{A,\v m,\v n,\v Q}^2(\h g,B,Y)=2(A+\b A) + \demi\left(\v p_L^2+\v p_R^2\right)$ are the masses of the boson/fermion pairs of superpartners. Integrating over $\tau_2$, the above expression for $Z_h$ leads to the free energy density (\ref{freegene}), while for $D=8$ the mass spectrum takes the more explicit form (\ref{stab7}).



\end{document}